\documentclass[english]{article}
\usepackage[T1]{fontenc}
\usepackage[latin9]{inputenc}
\usepackage{xcolor}
\usepackage{float}
\usepackage{amsmath}
\usepackage{amssymb}
\usepackage{cancel}

\makeatletter
\pdfoutput=1
\usepackage[T1]{fontenc}
\usepackage[latin9]{inputenc}
\usepackage{color}
\usepackage{float}
\usepackage{graphicx}
\usepackage{graphics}
\usepackage{esint}
\usepackage{enumerate}

\makeatletter

\usepackage{amsfonts,amssymb}
\usepackage{amsmath}
\usepackage{latexsym}
\usepackage{epsfig}
\usepackage{cite}
\usepackage{graphicx}
\usepackage[colorlinks,linkcolor=blue]{hyperref}
\newcommand{\be}{\begin{equation}}
\newcommand{\ee}{\end{equation}}

\allowdisplaybreaks

\usepackage{babel}
\usepackage{tcolorbox}

\usepackage{tikz}

\makeatother

\usepackage{babel}
\begin{document}
{}~ \hfill\vbox{\hbox{hep-th/yymmnnn}}\break
\vskip 3.0cm
\centerline{\Large \bf Black hole solutions in double field theory}
\vspace*{10.0ex}

\vspace*{10.0ex}
\centerline{\large Shuxuan Ying}
\vspace*{7.0ex}
\vspace*{4.0ex}

\centerline{\large \it Department of Physics}
\centerline{\large \it Chongqing University}
\centerline{\large \it Chongqing, 401331, China} \vspace*{1.0ex}

\vspace*{4.0ex}
\centerline{\large  ysxuan@cqu.edu.cn}

\vspace*{4.0ex}

\centerline{\bf Abstract} \bigskip \smallskip
In this paper, we study black hole solutions in double field theory. In the first part, we introduce a solution-generating method and classify black hole solutions into three categories in standard double field theory. To solve the problem of double time, we utilize space/time split double field theory in the second part to derive black hole solutions. By introducing a cosmological constant and imposing the strong constraint, our findings indicate that when considering the entire doubled spacetime, the curvature signifies a generalized AdS vacuum. However, in the subspace of the entire spacetime, black holes and curvature singularities emerge.

\vfill \eject
\baselineskip=16pt
\vspace*{10.0ex}
\tableofcontents

\section{Introduction}

Double field theory (DFT) is a recent progress in closed string theory.
This approach doubles all spacetime coordinates and elevates T-duality
to an intrinsic symmetry of the action \cite{Hull:2009mi,Hull:2009zb,Hohm:2010jy,Hohm:2010pp,Zwiebach:2011rg}.
In closed string theory, when a string moves in compactified spacetime,
the momentum of its center of mass can be quantized and is conjugate
to the coordinates $x^{i}$. Additionally, if the string wraps around
the compactified background multiple times, it introduces another
quantum number, the winding number. Consequently, the physical observables
of the string depend on both the momentum and the winding number.
T-duality, which exchanges momentum and winding numbers between two
strings wrapped on different compactified radii, requires treating
these two quantum numbers equally within a field theory. Thus, it
is reasonable to introduce extra coordinates $\tilde{x}_{i}$ conjugate
to the winding number. DFT suggests that the usual spacetime coordinates
$x^{i}\left(i=1,2,\cdots,D\right)$ can be doubled to $X^{M}=\left(\tilde{x}_{\mu},x^{\mu}\right)$,
where $M=1,2,\cdots,2D$. Consequently, all spacetime component fields,
including the gravitational field $g_{\mu\nu}$, the dilaton $\phi$,
and the anti-symmetric Kalb-Ramond field $b_{\mu\nu}$, depend on
$X^{M}$. In this framework, $g_{\mu\nu}$ and $b_{\mu\nu}$ combine
into a generalized metric that transforms linearly under continuous
$O\left(D,D,\mathbb{R}\right)$ transformations. The generalized coordinate
transformations are also well-defined on this $2D$ manifold. Based
on these features, the DFT action can be constructed to explicitly
exhibit $O\left(D,D,\mathbb{R}\right)$ symmetry. Moreover, DFT must
satisfy the strong constraint, meaning the theory lives in a $D$-dimensional
subspace of the $2D$ manifold. This constraint, derived from the
level-matching condition of closed string theory, requires that all
fields depend on only one set of the double coordinates $\left(\tilde{x},x\right)$.

In previous works \cite{Wu:2013sha,Wu:2013ixa,Lv:2014ava,Brandenberger:2017umf,Brandenberger:2018xwl,Brandenberger:2018bdc,Bernardo:2019pnq},
researchers have explored the cosmological solutions of DFT. These
studies have provided a consistent method to group T-dual cosmological
solutions from standard string cosmology \cite{Gasperini,Gasperini:2003pb,Gasperini:2004ss}.
These solutions unify pre- and post-big bang evolutions in a single
line element. Additionally, recent investigations have examined specific
black hole solutions under certain constraints \cite{Ko:2016dxa,Angus:2018mep,Liu:2021xfs,Li:2024ijj}.
However, two key issues remain unresolved:
\begin{enumerate}
\item The presence of two time coordinates in the generalized metric lacks
a reasonable physical explanation. Directly imposing the strong constraint
eliminates the double coordinates of the solutions, reducing them
to the standard string theory solutions.
\item The solutions appear to unify previous T-dual solutions without introducing
new physics.
\end{enumerate}
Solving these two problems could make studying black hole solutions
in DFT highly valuable. To tackle the first issue, we can utilize
recent advances in the canonical formulation of DFT \cite{Hohm:2013nja,Naseer:2015fba,Hohm:2022pfi}.
In this formulation, the ADM (Arnowitt-Deser-Misner)-like decomposition
splits the $2D$ manifold into temporal and spatial parts $X^{M}=\left(\tilde{t},t,X^{\hat{M}}\right)$,
requiring all spacetime component fields to be independent of the
dual time coordinate $\tilde{t}$. Consequently, only the time coordinate
remains undoubled, reducing the $2D$-dimensional DFT to $2D-1$ dimensions.
We refer to this approach as space/time split DFT for the remainder
of this paper. To illustrate how this decomposition can generate new
solutions, we must clarify the differences among three types of $O\left(D,D,\mathbb{R}\right)$
invariant theories as follows:
\begin{itemize}
\item \textbf{Low-energy effective action with $O\left(m,m,\mathbb{R}\right)$
duality:} In this action, spacetime coordinates are not doubled. The
$O\left(m,m,\mathbb{R}\right)$ dduality imposes significant constraints
on the form of the spacetime metric. Specifically, when the background
metric is independent of $m$ coordinates, the action exhibits $O\left(m,m,\mathbb{R}\right)$
symmetry. For example, FLRW backgrounds display $O\left(D-1,D-1,\mathbb{R}\right)$
symmetry, also known as \textquotedbl scale-factor duality,\textquotedbl{}
where the action remains invariant under the scale factor transformation
$a\left(t\right)\rightarrow\frac{1}{a\left(t\right)}$. Recent studies
have explored the classification of all orders of $\alpha^{\prime}$
corrections based on this duality \cite{Hohm:2015doa,Hohm:2019ccp,Hohm:2019jgu}.
Non-perturbative and non-singular solutions can be found in references\cite{Wang:2019mwi,Wang:2019kez,Wang:2019dcj,Wang:2020eln,Ying:2021xse,Ying:2022xaj,Ying:2022cix,Wu:2024eci}.
Solutions involving matter sources are discussed in references \cite{Bernardo:2019bkz,Bernardo:2020zlc,Bernardo:2020nol,Bernardo:2021xtr}.
\item \textbf{DFT with $O\left(D,D,\mathbb{R}\right)$ symmetry:} In this
formulation, all spacetime coordinates are doubled, and massless closed
string fields depend on these double coordinates without needing to
be independent of $m$ coordinates. The level-matching condition of
closed strings imposes the strong constraint, requiring the theory
to live on a $D-1$ subspace of the doubled spacetime. Vacuum cosmological
solutions of DFT have been provided in \cite{Wu:2013sha,Wu:2013ixa,Lv:2014ava}
and later extended to include matter sources \cite{Brandenberger:2017umf,Brandenberger:2018xwl,Brandenberger:2018bdc,Bernardo:2019pnq}.
Additionally, interesting solutions such as pp-waves \cite{Berkeley:2014nza},
monopoles \cite{Berman:2014jsa}, and a summary of these solutions
can be found in \cite{Berman:2020tqn}.
\item \textbf{Space/time split DFT with $O\left(D-1,D-1,\mathbb{R}\right)$
symmetry:} In this version, only the spatial coordinates are doubled,
and fields are not required to be independent of $m$ coordinates.
Since a metric component in the time direction is absent, solutions
benefit from the double coordinates while also extending beyond simple
combinations. This approach holds potential for discovering new solutions.
There have been some efforts to study black hole thermodynamics in
space/time split DFT \cite{Arvanitakis:2016zes} and cosmology \cite{Hohm:2022pfi,Lescano:2021nju}.
\end{itemize}
Moreover, it is important to note that compact spacetime is not required
from the perspective of field theory. For example, in the low-energy
effective action with $O\left(D,D,\mathbb{R}\right)$ duality, the
spacetime coordinates are not doubled. If the background is compactified,
even in field theory, the continuous $O\left(D,D,\mathbb{R}\right)$
symmetry becomes a discrete $O\left(D,D,\mathbb{Z}\right)$ symmetry.
Throughout the rest of this paper, we will abbreviate $O\left(D,D,\mathbb{R}\right)$
as $O\left(D,D\right)$.

The aim of this paper is to calculate the black hole solutions of
DFT. To preserve more non-trivial properties from $O\left(D,D\right)$
symmetry, we adhere to the following calculation principles: 1) Since
the strong constraint erases all features of DFT, we initially calculate
solutions that violate this constraint. Although these solutions currently
lack clear physical interpretations, we hope future studies will solve
this issue. 2) When imposing the strong constraint to obtain physical
solutions, we assume the generalized metric $\mathcal{H}_{MN}$ represents
the entire spacetime metric, not just the usual metric $g_{ij}$.
Based on these principles, we first calculate the black hole solutions
of standard DFT, which includes double time coordinates. These solutions
can be derived using the following three methods. Let us recall the
traditional low-energy effective action with a vanishing Kalb-Ramond
field for simplicity:

\begin{equation}
S=\int d^{D}x\sqrt{-g}e^{-2\phi}\left[R+4\left(\nabla\phi\right)^{2}\right].\label{eq:effective action 0}
\end{equation}

\noindent where $g_{ij}$ denotes the string metric and $\phi$ is
the dilaton. The black hole solutions of DFT can be constructed through
the solutions of this action (\ref{eq:effective action 0}):

{\noindent} \rule[-1pt]{16.5cm}{0.05em}

\vspace*{1.0ex}

\noindent \textbf{1st kind of solutions:} If $g_{ij}\left(x\right)$
and $\phi\left(x\right)$ are solutions of the effective action (\ref{eq:effective action 0}),
the solutions of DFT that satisfy the strong constraint can be given
as:

\noindent
\begin{equation}
ds^{2}=g^{ij}\left(x\right)d\tilde{x}_{i}d\tilde{x}_{j}+g_{ij}\left(x\right)dx^{i}dx^{j},\qquad or\qquad ds^{2}=\left(g^{-1}\right)^{ij}\left(\tilde{x}\right)d\tilde{x}_{i}d\tilde{x}_{j}+\left(g^{-1}\right)_{ij}\left(\tilde{x}\right)dx^{i}dx^{j}.
\end{equation}

\vspace*{1.0ex}

\noindent \textbf{2nd kind of solutions: }If $g_{ij}\left(x\right)$
and $\left(g^{-1}\right)_{ij}\left(x\right)$ are solutions of the
effective action (\ref{eq:effective action 0}), the constraint-satisfying
solutions of DFT can be given as\footnote{This kind of solutions are also discussed in Refs. \cite{Berkeley:2014nza,Berman:2014jsa}.}:

\begin{equation}
ds^{2}=g^{ij}\left(x\right)d\tilde{x}_{i}d\tilde{x}_{j}+g_{ij}\left(x\right)dx^{i}dx^{j},\qquad or\qquad ds^{2}=\left(g^{-1}\right)^{ij}\left(x\right)d\tilde{x}_{i}d\tilde{x}_{j}+\left(g^{-1}\right)_{ij}\left(x\right)dx^{i}dx^{j},
\end{equation}

\noindent or

\begin{equation}
ds^{2}=g^{ij}\left(\tilde{x}\right)d\tilde{x}_{i}d\tilde{x}_{j}+g_{ij}\left(\tilde{x}\right)dx^{i}dx^{j},\qquad or\qquad ds^{2}=\left(g^{-1}\right)^{ij}\left(\tilde{x}\right)d\tilde{x}_{i}d\tilde{x}_{j}+\left(g^{-1}\right)_{ij}\left(\tilde{x}\right)dx^{i}dx^{j}.
\end{equation}

\vspace*{1.0ex}

\noindent \textbf{3rd kind of solutions: }If $g_{ij}\left(x\right)$
and $\left(g^{-1}\right)_{ij}\left(x\right)$ are solutions of the
effective action (\ref{eq:effective action 0}), we can construct
constraint-violating solutions of DFT such that $g\left(\tilde{x},x\right)=g\left(\tilde{x}\right)\times\left(g^{-1}\right)\left(x\right)$.
These solutions depend on double coordinates simultaneously and take
the forms:

\begin{equation}
ds^{2}=g^{ij}\left(x,\tilde{x}\right)d\tilde{x}_{i}d\tilde{x}_{j}+g_{ij}\left(x,\tilde{x}\right)dx^{i}dx^{j},\qquad or\qquad ds^{2}=\left(g^{-1}\right)^{ij}\left(x,\tilde{x}\right)d\tilde{x}_{i}d\tilde{x}_{j}+\left(g^{-1}\right)_{ij}\left(x,\tilde{x}\right)dx^{i}dx^{j}.
\end{equation}

{\noindent} \rule[-1pt]{16.5cm}{0.05em}

\vspace*{2.0ex}

The problem with these solutions is the presence of double timelike
directions $dt$ and $d\tilde{t}$, which lack explicit physical interpretations.
To solve this issue, we consider the $2D-1$ dimensional space/time
split DFT, which is independent of the dual time coordinate. To obtain
non-trivial solutions, we consider non-critical strings, which introduce
the cosmological constant. This additional term is valid as the DFT
action should reduce to the low-energy effective action after imposing
the strong constraint \cite{Wu:2013sha,Lv:2014ava}. We then calculate
the equations of motion (EOM) for the corresponding action and use
the generated methods mentioned above to derive the black hole solutions.
The new insights of these solutions arise from the curvature definition
in doubled spacetime. In DFT, the Ricci curvature is generalized to
include the doubled coordinates and dilatonic dynamics, known as the
generalized Ricci scalar $\mathcal{R}$. It satisfies the EOM for
the dilaton and always vanishes ($\mathcal{R}=0$). Since we consider
the cosmological constant, the generalized Ricci scalar for the whole
doubled spacetime is:

\begin{equation}
\mathcal{R}=-\lambda^{2}.
\end{equation}

\noindent This implies that, regardless of the obtained solution,
the entire doubled spacetime is a generalized AdS vacuum. On the other
hand, since the geometry of the subspace $x$ of the doubled spacetime
can be completely determined by the low-energy effective action, we
can construct the black hole solution in this subspace. In other words,
considering the entire spacetime results in the generalized AdS vacuum,
whereas black holes and curvature singularities emerge in the subspace.
\emph{Finally, we note that a complete understanding of DFT black
holes is still far from being achieved. The primary challenge is the
lack of evidence confirming whether the generalized metric can be
interpreted as the physical spacetime of our universe. Therefore,
our focus is on obtaining solutions to the DFT action and providing
potential interpretations.}

The remainder of this paper is outlined as follows. In Section 2,
we review the dilatonic and gravitational EOM of standard DFT, classify
the solutions into three kinds, and present the main examples. In
Section 3, we introduce the space/time split DFT and calculate the
black hole solutions of its EOM. Section 4 contains our discussion
and conclusion.

\section{Standard double field theory}

\subsection{Equations of motion for the generalized metric and dilaton}

The EOM for the dilatonic and gravitational fields in DFT have been
extensively reviewed in the reference \cite{Hohm:2010pp}. Here, we
briefly summarize these equations. To construct the spacetime action
of DFT, we introduce the $2D\times2D$ ``generalized metric,'' which
incorporates the gravitational field $g_{ij}$ and the antisymmetric
Kalb-Ramond field $b_{ij}$:

\begin{equation}
\mathcal{H}_{MN}=\left(\begin{array}{cc}
g^{ij} & -g^{ik}b_{kj}\\
b_{ik}g^{kj} & g_{ij}-b_{ik}g^{kl}b_{lj}
\end{array}\right),\quad\mathcal{H}^{MN}=\left(\begin{array}{cc}
g_{ij}-b_{ik}g^{kl}b_{lj} & b_{ik}g^{kj}\\
-g^{ik}b_{kj} & g^{ij}
\end{array}\right),
\end{equation}

\noindent where $\mathcal{H}^{MP}\mathcal{H}_{PN}=\delta_{\;N}^{M}$,
$\mathcal{H}^{MN}\equiv\eta^{MP}\mathcal{H}_{PQ}\eta^{QN}=\mathcal{H}_{PQ}\eta^{PM}\eta^{QN}$,
and $\mathcal{H}^{MN}=\mathcal{H}^{NM}$ ($M,N,\ldots=1,2,\ldots,2D$).
The indices can be raised and lowered by the $O\left(D,D\right)$-invariant
metric:

\begin{equation}
\eta_{MN}=\left(\begin{array}{cc}
0 & I\\
I & 0
\end{array}\right),\qquad\eta^{MN}=\eta_{MN}.
\end{equation}

\noindent The gauge-invariant DFT action can thus be constructed from
$O\left(D,D\right)$ scalars:
\begin{eqnarray}
S & = & \int d^{D}xd^{D}\tilde{x}\mathcal{L}_{\mathrm{DFT}}\left(d,\mathcal{H}_{MN}\right)\nonumber \\
 & = & \int d^{D}xd^{D}\tilde{x}e^{-2d}\left(\frac{1}{8}\mathcal{H}^{MN}\partial_{M}\mathcal{H}^{KL}\partial_{N}\mathcal{H}_{KL}-\frac{1}{2}\mathcal{H}^{MN}\partial_{N}\mathcal{H}^{KL}\partial_{L}\mathcal{H}_{MK}\right.\nonumber \\
 &  & \left.-2\partial_{M}d\partial_{N}\mathcal{H}^{MN}+4\mathcal{H}^{MN}\partial_{M}d\partial_{N}d\right),\label{eq:DFT action}
\end{eqnarray}

\noindent where the derivatives are $\partial_{M}=\left(\tilde{\partial}^{i},\partial_{i}\right)$
and the $O\left(D,D\right)$-invariant dilaton $d$ is defined in
terms of the physical dilaton $\phi$ as:

\begin{equation}
e^{-2d}=\sqrt{g}e^{-2\phi}.
\end{equation}

\noindent Since each term in the action is a scalar, the $O\left(D,D\right)$
symmetry is manifest. Moreover, the action (\ref{eq:DFT action})
can also be rewritten in terms of the generalized scalar curvature
$\mathcal{R}\left(d,\mathcal{H}_{MN}\right)$ with some additional
boundary terms:

\begin{equation}
\mathcal{L}_{\mathrm{DFT}}\left(d,\mathcal{H}_{MN}\right)=e^{-2d}\mathcal{R}\left(d,\mathcal{H}_{MN}\right)+\partial_{M}\left(e^{-2d}\left[\partial_{N}\mathcal{H}^{MN}-4\mathcal{H}^{MN}\partial_{N}d\right]\right),
\end{equation}

\noindent where the generalized curvature scalar is defined as:

\begin{eqnarray}
\mathcal{R}\left(d,\mathcal{H}_{MN}\right) & = & \frac{1}{8}\mathcal{H}^{MN}\partial_{M}\mathcal{H}^{KL}\partial_{N}\mathcal{H}_{KL}-\frac{1}{2}\mathcal{H}^{MN}\partial_{N}\mathcal{H}^{KL}\partial_{L}\mathcal{H}_{MK}+4\mathcal{H}^{MN}\partial_{M}\partial_{N}d\nonumber \\
 &  & +4\partial_{M}\mathcal{H}^{MN}\partial_{N}d-4\mathcal{H}^{MN}\partial_{M}d\partial_{N}d-\partial_{M}\partial_{N}\mathcal{H}^{MN}.
\end{eqnarray}

\noindent Additionally, all fields must satisfy the strong constraint
$\eta^{MN}\partial_{M}\partial_{N}f=0$ (or equivalently $\partial^{M}f\partial_{M}g=0$)due
to the level-matching condition of closed string theory. This constraint
requires all fields to live only on a subspace of the doubled coordinates.
The gauge transformations of the DFT action (\ref{eq:DFT action})
are given by:

\begin{eqnarray}
\delta_{\xi}\mathcal{H}^{MN} & = & \hat{\mathcal{L}}_{\xi}\mathcal{H}^{MN}\equiv\xi^{P}\partial_{P}\mathcal{H}^{MN}+\left(\partial^{M}\xi_{P}-\partial_{P}\xi^{M}\right)\mathcal{H}^{PN}+\left(\partial^{N}\xi_{P}-\partial_{P}\xi^{N}\right)\mathcal{H}^{MP},\nonumber \\
\delta d & = & \xi^{M}\partial_{M}d-\frac{1}{2}\partial_{M}\xi^{M},
\end{eqnarray}

\noindent with $\xi^{M}=\left(\hat{\xi}_{i},\xi^{i}\right)$ and ``generalized
Lie derivatives'' $\hat{\mathcal{L}}_{\xi}$. The dilatonic EOM can
be derived from the variation of the DFT action: (\ref{eq:DFT action}):

\begin{eqnarray}
\delta_{d}S & = & \int dxd\tilde{x}e^{-2d}\left(-\frac{1}{4}\mathcal{H}^{MN}\partial_{M}\mathcal{H}^{KL}\partial_{N}\mathcal{H}_{KL}+\mathcal{H}^{MN}\partial_{N}\mathcal{H}^{KL}\partial_{L}\mathcal{H}_{MK}\right.\nonumber \\
 &  & \left.+2\partial_{M}\partial_{N}\mathcal{H}^{MN}+8\mathcal{H}^{MN}\partial_{M}d\partial_{N}d-8\partial_{M}\mathcal{H}^{MN}\partial_{N}d-8\mathcal{H}^{MN}\partial_{M}\partial_{N}d\right)\delta d.
\end{eqnarray}

\noindent This gives:

\begin{eqnarray}
 &  & \frac{1}{8}\mathcal{H}^{MN}\partial_{M}\mathcal{H}^{KL}\partial_{N}\mathcal{H}_{KL}-\frac{1}{2}\mathcal{H}^{MN}\partial_{M}\mathcal{H}^{KL}\partial_{K}\mathcal{H}_{NL}\nonumber \\
 &  & -\partial_{M}\partial_{N}\mathcal{H}^{MN}-4\mathcal{H}^{MN}\partial_{M}d\partial_{N}d+4\partial_{M}\mathcal{H}^{MN}\partial_{N}d+4\mathcal{H}^{MN}\partial_{M}\partial_{N}d=0.
\end{eqnarray}

\noindent To obtain the gravitational field equation for the generalized
metric $\mathcal{H}$, we consider the variation of $\mathcal{H}^{MN}$:

\begin{equation}
\delta_{\mathcal{H}}S=\int dxd\tilde{x}e^{-2d}\delta\mathcal{H}^{MN}\mathcal{K}_{MN},\label{eq:vary of action}
\end{equation}

\noindent where

\begin{eqnarray}
\mathcal{K}_{MN} & \equiv & \frac{1}{8}\partial_{M}\mathcal{H}^{KL}\partial_{N}\mathcal{H}_{KL}-\frac{1}{4}\left(\partial_{L}-2\partial_{L}d\right)\left(\mathcal{H}^{LK}\partial_{K}\mathcal{H}_{MN}\right)+2\partial_{M}\partial_{N}d\nonumber \\
 &  & -\frac{1}{2}\partial_{\left(N\right.}\mathcal{H}^{KL}\partial_{L}\mathcal{H}_{\left.M\right)K}+\frac{1}{2}\left(\partial_{L}-2\partial_{L}d\right)\left(\mathcal{H}^{KL}\partial_{\left(N\right.}\mathcal{H}_{\left.M\right)K}+\mathcal{H}_{\quad\left(M\right.}^{K}\partial_{K}\mathcal{H}_{\quad\left.N\right)}^{L}\right).\label{eq:K}
\end{eqnarray}

\noindent However, $\mathcal{K}_{MN}=0$ is not the gravitational
field equation, as this equation is not an $O\left(D,D\right)$-invariant
field equation. To obtain the $O\left(D,D\right)$-invariant EOM,
we first utilize the notations $\mathcal{H}\equiv\mathcal{H}^{\bullet\bullet}$
and $\eta\equiv\mathcal{\eta}_{\bullet\bullet}$. Since the $O\left(D,D\right)$
symmetry requires $\mathcal{H}$ to satisfy $\mathcal{H}\eta\mathcal{H}=\eta^{-1}$,
its varied field $\mathcal{H}^{\prime}=\mathcal{H}+\delta\mathcal{H}$
also needs to satisfy the symmetric condition $\mathcal{H}^{\prime}\eta\mathcal{H}^{\prime}=\eta^{-1}$.
Then we have:

\begin{equation}
\delta\mathcal{H}\eta\mathcal{H}+\mathcal{H}\eta\delta\mathcal{H}=0.
\end{equation}

\noindent To simplify our calculations, we define the following:

\begin{equation}
S_{\quad N}^{M}\equiv\mathcal{H}_{\quad N}^{M}=\eta^{MP}\mathcal{H}_{PN}=\mathcal{H}^{MP}\eta_{PN},\quad S^{2}=1.
\end{equation}

\noindent Using the notation $S\equiv S_{\quad\bullet}^{\bullet}=\mathcal{H}\eta$,
we derive the variation condition for

\begin{equation}
\delta\mathcal{H}S^{t}+S\delta\mathcal{H}=0.
\end{equation}

\noindent Given that $S^{2}=1$, we can rewrite the variation as:

\begin{equation}
\delta\mathcal{H}=-S\delta\mathcal{H}S^{t}.
\end{equation}

\noindent This can be further expressed as:

\begin{equation}
\delta\mathcal{H}=\frac{1}{4}\left(1+S\right)\mathcal{M}\left(1-S^{t}\right)+\frac{1}{4}\left(1-S\right)\mathcal{M}\left(1+S^{t}\right),
\end{equation}
where $\mathcal{M}$ is an arbitrary symmetric matrix ensuring the
symmetry of $\delta\mathcal{H}$. Substituting this back into the
variation of the action (\ref{eq:vary of action}), we get:

\begin{equation}
\frac{1}{4}\left(1-S^{t}\right)\mathcal{K}\left(1+S\right)+\frac{1}{4}\left(1+S^{t}\right)\mathcal{K}\left(1-S\right)=0.
\end{equation}

\noindent Thus, the $O\left(D,D\right)$-invariant gravitational field
equation is obtained:

\begin{eqnarray}
\mathcal{R}_{MN} & = & \frac{1}{4}\left(\delta_{M}^{\quad P}-S_{\quad M}^{P}\right)\mathcal{K}_{PQ}\left(\delta_{\quad N}^{Q}+S_{\quad N}^{Q}\right)+\frac{1}{4}\left(\delta_{M}^{\quad P}+S_{\quad M}^{P}\right)\mathcal{K}_{PQ}\left(\delta_{\quad N}^{Q}-S_{\quad N}^{Q}\right)\nonumber \\
 & = & \frac{1}{2}\mathcal{K}_{MN}-\frac{1}{2}S_{\quad M}^{P}\mathcal{K}_{PQ}S_{\quad N}^{Q}.
\end{eqnarray}

\noindent This is a T-duality covariant generalized tensor, which
serves as a natural extension of the Ricci tensors in Riemannian geometry.\textcolor{red}{{}
}In summary, the dilatonic and gravitational EOM of DFT are presented
as follows:

\begin{eqnarray}
\frac{1}{8}\mathcal{H}^{MN}\partial_{M}\mathcal{H}^{KL}\partial_{N}\mathcal{H}_{KL}-\frac{1}{2}\mathcal{H}^{MN}\partial_{M}\mathcal{H}^{KL}\partial_{K}\mathcal{H}_{NL}-\partial_{M}\partial_{N}\mathcal{H}^{MN}\nonumber \\
-4\mathcal{H}^{MN}\partial_{M}d\partial_{N}d+4\partial_{M}\mathcal{H}^{MN}\partial_{N}d+4\mathcal{H}^{MN}\partial_{M}\partial_{N}d & = & 0,\nonumber \\
\mathcal{K}_{MN}-S_{\quad M}^{P}\mathcal{K}_{PQ}S_{\quad N}^{Q} & = & 0.\label{eq:EOM}
\end{eqnarray}

\subsection{Some notations and definitions}

To calculate the EOM (\ref{eq:EOM}) in DFT, we need to define several
notations and calculation rules. First, we clarify the notations for
the generalized metric. The spatial vector in DFT is defined as:

\begin{equation}
\partial_{M}\equiv\left(\begin{array}{c}
\tilde{\partial}^{i}\\
\partial_{i}
\end{array}\right),
\end{equation}

\noindent where $i$ denotes the $D-1$ dimensional spatial coordinates.
The corresponding dual vector is given by:

\begin{equation}
dX^{M}\equiv\left(\begin{array}{c}
d\tilde{x}_{i}\\
dx^{i}
\end{array}\right).
\end{equation}

\noindent Using the dual vector, we can define the generalized line
element in terms of the generalized metric:

\begin{equation}
ds^{2}=\mathcal{H}_{MN}dX^{M}dX^{N}.
\end{equation}

\noindent For simplicity, we use block matrix notation to rewrite
the vector, the dual vector, and the generalized metric:

\begin{equation}
\partial_{M}=\left(\begin{array}{c}
\partial_{\mathbf{1}}\\
\partial_{\mathbf{2}}
\end{array}\right),\quad dX^{M}=\left(\begin{array}{c}
dX^{\mathbf{\mathbf{1}}}\\
dX^{\mathbf{2}}
\end{array}\right),\quad\mathcal{H}_{MN}=\left(\begin{array}{cc}
\mathcal{H}_{\mathbf{1}\mathbf{1}} & \mathcal{H}_{\mathbf{1}\mathbf{2}}\\
\mathcal{H}_{\mathbf{2}\mathbf{1}} & \mathcal{H}_{\mathbf{2}\mathbf{2}}
\end{array}\right),\quad\mathcal{H}^{MN}=\left(\begin{array}{cc}
\mathcal{H}^{\mathbf{1}\mathbf{1}} & \mathcal{H}^{\mathbf{1}\mathbf{2}}\\
\mathcal{H}^{\mathbf{2}\mathbf{1}} & \mathcal{H}^{\mathbf{2}\mathbf{2}}
\end{array}\right),
\end{equation}

\noindent where $\mathbf{1}$ represents the set of coordinates $\tilde{x}_{i}$
and $\mathbf{2}$ denotes the coordinates $x^{i}$.Using these notations,
the components of the generalized metric can be divided into four
blocks:

\begin{eqnarray}
\mathcal{H}_{\mathbf{1}\mathbf{1}}\left(\frac{\partial}{\partial\tilde{x}_{i}},\frac{\partial}{\partial\tilde{x}_{j}}\right)=g^{ij}, &  & \mathcal{H}_{\mathbf{1}\mathbf{2}}\left(\frac{\partial}{\partial\tilde{x}_{i}},\frac{\partial}{\partial x^{j}}\right)=-g^{ik}b_{kj},\nonumber \\
\mathcal{H}_{\mathbf{2}\mathbf{1}}\left(\frac{\partial}{\partial x^{i}},\frac{\partial}{\partial\tilde{x}_{j}}\right)=b_{ik}g^{kj}, &  & \mathcal{H}_{\mathbf{2}\mathbf{2}}\left(\frac{\partial}{\partial x^{i}},\frac{\partial}{\partial x^{j}}\right)=g_{ij}-b_{ik}g^{kl}b_{lj}.\label{eq:metric block}
\end{eqnarray}

\noindent It is worth noting that $g\left(x,\tilde{x}\right)$, $b\left(x,\tilde{x}\right)$,
and the dilaton $d\left(x,\tilde{x}\right)$ depend on two sets of
coordinates. The generalized line element becomes:

\begin{eqnarray}
dS^{2} & = & \mathcal{H}_{\mathbf{1}\mathbf{1}}dX^{\mathbf{1}}dX^{\mathbf{1}}+\mathcal{H}_{\mathbf{1}\mathbf{2}}dX^{\mathbf{1}}dX^{\mathbf{2}}+\mathcal{H}_{\mathbf{2}\mathbf{1}}dX^{\mathbf{2}}dX^{\mathbf{1}}+\mathcal{H}_{\mathbf{2}\mathbf{2}}dX^{\mathbf{2}}dX^{\mathbf{2}}\nonumber \\
 & = & g^{ij}d\tilde{x}_{i}d\tilde{x}_{j}-g^{ik}b_{kj}d\tilde{x}_{i}dx^{j}+b_{ik}g^{kj}dx^{i}d\tilde{x}_{j}+\left(g_{ij}-b_{ik}g^{kl}b_{lj}\right)dx^{i}dx^{j}.
\end{eqnarray}

\noindent For simplicity in the rest of the paper, we will set the
Kalb-Ramond field to be zero, so the line element becomes:

\begin{equation}
dS^{2}=\mathcal{H}_{\mathbf{1}\mathbf{1}}dX^{\mathbf{1}}dX^{\mathbf{1}}+\mathcal{H}_{\mathbf{2}\mathbf{2}}dX^{\mathbf{2}}dX^{\mathbf{2}}=g^{ij}d\tilde{x}_{i}d\tilde{x}_{j}+g_{ij}dx^{i}dx^{j}.
\end{equation}

\noindent Then, the blocks of the generalized metric are:

\begin{equation}
\mathcal{H}_{\mathbf{1}\mathbf{1}}=g^{ij},\qquad\mathcal{H}^{\mathbf{1}\mathbf{1}}=g_{ij},\qquad\mathcal{H}^{\mathbf{2}\mathbf{2}}=g^{ij},\qquad\mathcal{H}_{\mathbf{\mathbf{2}}\mathbf{2}}=g_{ij}.
\end{equation}

\noindent To calculate the contraction of $O\left(D,D\right)$ indices,
we introduce extra indices $\mathcal{H}^{\mathbf{1}\left(i\right)\mathbf{1}\left(j\right)}$
to represent the contractions of the elements of the block matrix.
Therefore, the generalized metric can be rewritten by adding the extra
indices:

\begin{equation}
\mathcal{H}_{M\left(i\right)N\left(j\right)}=\left(\begin{array}{cc}
\mathcal{H}_{\mathbf{1}\left(i\right)\mathbf{1}\left(j\right)} & 0\\
0 & \mathcal{H}_{\mathbf{2}\left(i\right)\mathbf{2}\left(j\right)}
\end{array}\right)=\left(\begin{array}{cc}
g^{ij} & 0\\
0 & g_{ij}
\end{array}\right),
\end{equation}

\noindent where $b_{ij}=0$. Now, there exist two sets of indices:
the block matrix notations $M$, $N$ and the indices of the components
of the block matrix $i$, $j$. Based on these new notations, we give
the following examples that we will use in the calculations:

\begin{eqnarray}
\mathcal{H}^{M\left(i\right)N\left(j\right)}\partial_{M\left(i\right)}d\partial_{N\left(j\right)}d & = & \mathcal{H}^{\mathbf{1}\left(i\right)\mathbf{1}\left(j\right)}\partial_{\mathbf{1}\left(i\right)}d\partial_{\mathbf{1}\left(j\right)}d+\mathcal{H}^{\mathbf{2}\left(i\right)\mathbf{2}\left(j\right)}\partial_{\mathbf{2}\left(i\right)}d\partial_{\mathbf{2}\left(j\right)}d\nonumber \\
 & = & g_{ij}\tilde{\partial}^{i}d\tilde{\partial}^{j}d+g^{ij}\partial_{i}d\partial_{j}d.
\end{eqnarray}

\noindent Moreover, for the derivatives, we have:

\begin{eqnarray}
\partial_{\mathbf{1}\left(k\right)}\mathcal{H}^{\mathbf{2}\left(i\right)\mathbf{2}\left(j\right)} & = & \tilde{\partial}^{k}g^{ij}\nonumber \\
\partial_{\mathbf{2}\left(k\right)}\mathcal{H}^{\mathbf{1}\left(i\right)\mathbf{1}\left(j\right)} & = & \partial_{k}g_{ij}\nonumber \\
\partial_{\mathbf{1}\left(p\right)}\mathcal{H}^{\mathbf{1}\left(i\right)\mathbf{1}\left(j\right)}\partial_{\mathbf{2}\left(q\right)}\mathcal{H}_{\mathbf{1}\left(i\right)\mathbf{1}\left(j\right)} & = & \tilde{\partial}^{p}g_{ij}\partial_{q}g^{ij}
\end{eqnarray}

\noindent On the other hand, the rules for $S_{\quad N}^{M}$ are

\begin{equation}
\mathcal{H}_{\quad\mathbf{1}\left(j\right)}^{\mathbf{2}\left(i\right)}=g^{ij},\quad\mathcal{H}_{\quad\mathbf{2}\left(j\right)}^{\mathbf{1}\left(i\right)}=g_{ij}.
\end{equation}

\noindent and

\begin{equation}
\partial_{\mathbf{2}\left(k\right)}\mathcal{H}_{\quad\mathbf{2}\left(j\right)}^{\mathbf{1}\left(i\right)}=\partial_{k}g_{ij},\qquad\partial_{\mathbf{1}\left(k\right)}\mathcal{H}_{\quad\mathbf{1}\left(j\right)}^{\mathbf{2}\left(i\right)}=\tilde{\partial}^{k}g^{ij}.
\end{equation}

\noindent Note that the raising and lowering of indices are given
by:

\begin{equation}
g_{ij}\tilde{\partial}^{i}=\tilde{\partial}_{j},\quad g^{ij}\partial_{i}=\partial^{j}.
\end{equation}

\subsection{Classification of black hole solutions}

In this subsection, our goal is to outline the methodology for generating
typical DFT black hole solutions. These solutions are categorized
into three types, each accompanied by specific examples. However,
due to the inclusion of dual time coordinates, the physical interpretation
of these solutions is not explicitly defined.

Firstly, let us revisit the EOM for the graviton and dilaton fields:

\begin{eqnarray}
\frac{1}{8}\mathcal{H}^{MN}\partial_{M}\mathcal{H}^{KL}\partial_{N}\mathcal{H}_{KL}-\frac{1}{2}\mathcal{H}^{MN}\partial_{M}\mathcal{H}^{KL}\partial_{K}\mathcal{H}_{NL}-\partial_{M}\partial_{N}\mathcal{H}^{MN}\nonumber \\
-4\mathcal{H}^{MN}\partial_{M}d\partial_{N}d+4\partial_{M}\mathcal{H}^{MN}\partial_{N}d+4\mathcal{H}^{MN}\partial_{M}\partial_{N}d & = & 0,\nonumber \\
2\mathcal{R}_{MN}=\mathcal{K}_{MN}-S_{\quad M}^{P}\mathcal{K}_{PQ}S_{\quad N}^{Q} & = & 0.
\end{eqnarray}

\noindent Assuming the Kalb-Ramond field is zero and substituting
the generalized metric formulation:

\begin{equation}
\mathcal{H}_{M\left(i\right)N\left(j\right)}=\left(\begin{array}{cc}
\mathcal{H}_{\mathbf{1}\left(i\right)\mathbf{1}\left(j\right)} & 0\\
0 & \mathcal{H}_{\mathbf{2}\left(i\right)\mathbf{2}\left(j\right)}
\end{array}\right)=\left(\begin{array}{cc}
g^{ij} & 0\\
0 & g_{ij}
\end{array}\right),
\end{equation}

\noindent the EOM can be expanded using the notation and rules established
in the previous section. Specifically for the graviton field, we obtain:

\begin{eqnarray}
\mathcal{R}_{\mathbf{1}\left(p\right)\mathbf{1}\left(q\right)} & = & \frac{1}{8}\tilde{\partial}^{p}g_{ij}\tilde{\partial}^{q}g^{ij}+\frac{1}{8}\tilde{\partial}^{p}g^{ij}\tilde{\partial}^{q}g_{ij}-\frac{1}{4}\tilde{\partial}^{i}\left(g_{ij}\tilde{\partial}^{j}g^{pq}\right)-\frac{1}{4}\partial_{i}\left(g^{ij}\partial_{j}g^{pq}\right)\nonumber \\
 &  & -\frac{1}{4}\tilde{\partial}^{q}g_{ij}\tilde{\partial}^{j}g^{pi}-\frac{1}{4}\tilde{\partial}^{p}g_{ij}\tilde{\partial}^{j}g^{qi}+\frac{1}{4}\tilde{\partial}^{i}\left(g_{ji}\tilde{\partial}^{q}g^{pj}\right)+\frac{1}{4}\tilde{\partial}^{i}\left(g_{ji}\tilde{\partial}^{p}g^{qj}\right)\nonumber \\
 &  & +\frac{1}{4}\partial_{i}\left(g^{jp}\partial_{j}g^{iq}\right)+\frac{1}{4}\partial_{i}\left(g^{jq}\partial_{j}g^{ip}\right)-\frac{1}{8}g^{ip}\partial_{i}g_{kl}\partial_{j}g^{kl}g^{jq}-\frac{1}{8}g^{ip}\partial_{i}g^{kl}\partial_{j}g_{kl}g^{jq}\nonumber \\
 &  & +\frac{1}{4}g^{ip}\tilde{\partial}^{l}\left(g_{lk}\tilde{\partial}^{k}g_{ij}\right)g^{jq}+\frac{1}{4}g^{ip}\partial_{l}\left(g^{lk}\partial_{k}g_{ij}\right)g^{jq}+\frac{1}{4}g^{ip}\partial_{j}g^{kl}\partial_{l}g_{ik}g^{jq}+\frac{1}{4}g^{ip}\partial_{i}g^{kl}\partial_{l}g_{jk}g^{jq}\nonumber \\
 &  & -\frac{1}{4}g^{ip}\partial_{l}\left(g^{kl}\partial_{j}g_{ik}\right)g^{jq}-\frac{1}{4}g^{ip}\partial_{l}\left(g^{kl}\partial_{i}g_{jk}\right)g^{jq}-\frac{1}{4}g^{ip}\tilde{\partial}^{l}\left(g_{ki}\tilde{\partial}^{k}g_{lj}\right)g^{jq}-\frac{1}{4}g^{ip}\tilde{\partial}^{l}\left(g_{kj}\tilde{\partial}^{k}g_{li}\right)g^{jq}\nonumber \\
 &  & +\frac{1}{2}\tilde{\partial}^{i}d\left(g_{ij}\tilde{\partial}^{j}g^{pq}\right)+\frac{1}{2}\partial_{i}d\left(g^{ij}\partial_{j}g^{pq}\right)+2\tilde{\partial}^{p}\tilde{\partial}^{q}d\nonumber \\
 &  & -\frac{1}{2}\partial_{i}d\left(g^{jp}\partial_{j}g^{iq}\right)-\frac{1}{2}\partial_{i}d\left(g^{jq}\partial_{j}g^{ip}\right)-\frac{1}{2}\tilde{\partial}^{i}d\left(g_{ji}\tilde{\partial}^{q}g^{pj}\right)-\frac{1}{2}\tilde{\partial}^{i}d\left(g_{ji}\tilde{\partial}^{p}g^{qj}\right)\nonumber \\
 &  & -\frac{1}{2}g^{ip}\tilde{\partial}^{l}d\left(g_{lk}\tilde{\partial}^{k}g_{ij}\right)g^{jq}-\frac{1}{2}g^{ip}\partial_{l}d\left(g^{lk}\partial_{k}g_{ij}\right)g^{jq}-2g^{ip}\partial_{i}\partial_{j}dg^{jq}\nonumber \\
 &  & +\frac{1}{2}g^{ip}\partial_{l}d\left(g^{kl}\partial_{j}g_{ik}\right)g^{jq}+\frac{1}{2}g^{ip}\partial_{l}d\left(g^{kl}\partial_{i}g_{jk}\right)g^{jq}\nonumber \\
 &  & +\frac{1}{2}g^{ip}\tilde{\partial}^{l}d\left(g_{ki}\tilde{\partial}^{k}g_{lj}\right)g^{jq}+\frac{1}{2}g^{ip}\tilde{\partial}^{l}d\left(g_{kj}\tilde{\partial}^{k}g_{li}\right)g^{jq}.\nonumber \\
\\
\mathcal{R}_{\mathbf{2}\left(p\right)\mathbf{2}\left(q\right)} & = & \frac{1}{8}\partial_{p}g_{ij}\partial_{q}g^{ij}+\frac{1}{8}\partial_{p}g^{ij}\partial_{q}g_{ij}-\frac{1}{4}\tilde{\partial}^{i}\left(g_{ij}\tilde{\partial}^{j}g_{pq}\right)-\frac{1}{4}\partial_{i}\left(g^{ij}\partial_{j}g_{pq}\right)\nonumber \\
 &  & -\frac{1}{4}\partial_{q}g^{ij}\partial_{j}g_{pi}-\frac{1}{4}\partial_{p}g^{ij}\partial_{j}g_{qi}+\frac{1}{4}\partial_{i}\left(g^{ji}\partial_{q}g_{pj}\right)+\frac{1}{4}\partial_{i}\left(g^{ji}\partial_{p}g_{qj}\right)\nonumber \\
 &  & +\frac{1}{4}\tilde{\partial}^{i}\left(g_{jp}\tilde{\partial}^{j}g_{iq}\right)+\frac{1}{4}\tilde{\partial}^{i}\left(g_{jq}\tilde{\partial}^{j}g_{ip}\right)-\frac{1}{8}g_{ip}\tilde{\partial}^{i}g_{kl}\tilde{\partial}^{j}g^{kl}g_{jq}-\frac{1}{8}g_{ip}\tilde{\partial}^{i}g^{kl}\tilde{\partial}^{j}g_{kl}g_{jq}\nonumber \\
 &  & +\frac{1}{4}g_{ip}\tilde{\partial}^{l}\left(g_{lk}\tilde{\partial}^{k}g^{ij}\right)g_{jq}+\frac{1}{4}g_{ip}\partial_{l}\left(g^{lk}\partial_{k}g^{ij}\right)g_{jq}+\frac{1}{4}g_{ip}\tilde{\partial}^{j}g_{kl}\tilde{\partial}^{l}g^{ik}g_{jq}+\frac{1}{4}g_{ip}\tilde{\partial}^{i}g_{kl}\tilde{\partial}^{l}g^{jk}g_{jq}\nonumber \\
 &  & -\frac{1}{4}g_{ip}\tilde{\partial}^{l}\left(g_{kl}\tilde{\partial}^{j}g^{ik}\right)g_{jq}-\frac{1}{4}g_{ip}\tilde{\partial}^{l}\left(g_{kl}\tilde{\partial}^{i}g\right)g_{jq}-\frac{1}{4}g_{ip}\partial_{l}\left(g^{ki}\partial_{k}g^{lj}\right)g_{jq}-\frac{1}{4}g_{ip}\partial_{l}\left(g^{kj}\partial_{k}g^{li}\right)g_{jq}\nonumber \\
 &  & +\frac{1}{2}\tilde{\partial}^{i}d\left(g_{ij}\tilde{\partial}^{j}g_{pq}\right)+\frac{1}{2}\partial_{i}d\left(g^{ij}\partial_{j}g_{pq}\right)+2\partial_{p}\partial_{q}d\nonumber \\
 &  & -\frac{1}{2}\tilde{\partial}^{i}d\left(g_{jp}\tilde{\partial}^{j}g_{iq}\right)-\frac{1}{2}\tilde{\partial}^{i}d\left(g_{jq}\tilde{\partial}^{j}g_{ip}\right)-\frac{1}{2}\partial_{i}d\left(g^{ji}\partial_{q}g_{pj}\right)-\frac{1}{2}\partial_{i}d\left(g^{ji}\partial_{p}g_{qj}\right)\nonumber \\
 &  & -\frac{1}{2}g_{ip}\tilde{\partial}^{l}d\left(g_{lk}\tilde{\partial}^{k}\tilde{g}^{ij}\right)g_{jq}-\frac{1}{2}g_{ip}\partial_{l}d\left(g^{lk}\partial_{k}g^{ij}\right)g_{jq}-2g_{ip}\tilde{\partial}^{i}\tilde{\partial}^{j}dg_{jq}\nonumber \\
 &  & +\frac{1}{2}g_{ip}\tilde{\partial}^{l}d\left(g_{kl}\tilde{\partial}^{j}g^{ik}\right)g_{jq}+\frac{1}{2}g_{ip}\tilde{\partial}^{l}d\left(g_{kl}\tilde{\partial}^{i}g^{jk}\right)g_{jq}\nonumber \\
 &  & +\frac{1}{2}g_{ip}\partial_{l}d\left(g^{ki}\partial_{k}g^{lj}\right)g_{jq}+\frac{1}{2}g_{ip}\partial_{l}d\left(g^{kj}\partial_{k}g^{li}\right)g_{jq},\nonumber \\
\\
\mathcal{R}_{\mathbf{1}\left(p\right)\mathbf{2}\left(q\right)} & = & \frac{1}{8}\tilde{\partial}^{p}g_{ij}\partial_{q}g^{ij}+\frac{1}{8}\tilde{\partial}^{p}g^{ij}\partial_{q}g_{ij}-\frac{1}{4}\partial_{q}g_{ij}\tilde{\partial}^{j}g^{pi}-\frac{1}{4}\tilde{\partial}^{p}g^{ij}\partial_{j}g_{qi}\nonumber \\
 &  & +\frac{1}{4}\tilde{\partial}^{i}\left(g_{ji}\partial_{q}g^{pj}\right)+\frac{1}{4}\partial_{i}\left(g^{ji}\tilde{\partial}^{p}g_{qj}\right)+\frac{1}{4}\tilde{\partial}^{i}\left(g^{jp}\partial_{j}g_{iq}\right)+\frac{1}{4}\partial_{i}\left(g_{jq}\tilde{\partial}^{j}g^{ip}\right)\nonumber \\
 &  & -\frac{1}{8}g^{ip}\partial_{i}g_{kl}\tilde{\partial}^{j}g^{kl}g_{jq}-\frac{1}{8}g^{ip}\partial_{i}g^{kl}\tilde{\partial}^{j}g_{kl}g_{jq}+\frac{1}{4}g^{ip}\partial_{i}g_{kl}\tilde{\partial}^{l}g^{jk}g_{jq}+\frac{1}{4}g^{ip}\tilde{\partial}^{j}g^{kl}\partial_{l}g_{ik}g_{jq}\nonumber \\
 &  & -\frac{1}{4}g^{ip}\tilde{\partial}^{l}\left(g_{kl}\partial_{i}g^{jk}\right)g_{jq}-\frac{1}{4}g^{ip}\partial_{l}\left(g^{kl}\tilde{\partial}^{j}g_{ik}\right)g_{jq}-\frac{1}{4}g^{ip}\partial_{l}\left(g_{ki}\tilde{\partial}^{k}g^{lj}\right)g_{jq}-\frac{1}{4}g^{ip}\tilde{\partial}^{l}\left(g^{kj}\partial_{k}g_{li}\right)g_{jq}\nonumber \\
 &  & +2\tilde{\partial}^{p}\partial_{q}d-\tilde{\partial}^{i}d\left(g^{jp}\partial_{j}g_{iq}\right)-\partial_{i}d\left(g_{jq}\tilde{\partial}^{j}g^{ip}\right)\nonumber \\
 &  & -\tilde{\partial}^{i}d\left(g_{ji}\partial_{q}g^{pj}\right)-\partial_{i}d\left(g^{ji}\tilde{\partial}^{p}g_{qj}\right)-2g^{ip}\partial_{i}\tilde{\partial}^{j}dg_{jq}\nonumber \\
 &  & +\frac{1}{2}g^{ip}\tilde{\partial}^{l}d\left(g_{kl}\partial_{i}g^{jk}\right)g_{jq}+\frac{1}{2}g^{ip}\partial_{l}d\left(g^{kl}\tilde{\partial}^{j}g_{ik}\right)g_{jq}\nonumber \\
 &  & +\frac{1}{2}g^{ip}\tilde{\partial}^{l}d\left(g^{kj}\partial_{k}g_{li}\right)g_{jq}+\frac{1}{2}g^{ip}\partial_{l}d\left(g_{ki}\tilde{\partial}^{k}g^{lj}\right)g_{jq},\nonumber \\
\\
\mathcal{R}_{\mathbf{2}\left(p\right)\mathbf{1}\left(q\right)} & = & \frac{1}{8}\partial_{p}g_{ij}\tilde{\partial}^{q}g^{ij}+\frac{1}{8}\partial_{p}g^{ij}\tilde{\partial}^{q}g_{ij}-\frac{1}{4}\partial_{p}g_{ij}\tilde{\partial}^{j}g^{qi}-\frac{1}{4}\tilde{\partial}^{q}g^{ij}\partial_{j}g_{pi}\nonumber \\
 &  & +\frac{1}{4}\tilde{\partial}^{i}\left(g_{ji}\partial_{p}g^{qj}\right)+\frac{1}{4}\partial_{i}\left(g^{ji}\tilde{\partial}^{q}g_{pj}\right)+\frac{1}{4}\tilde{\partial}^{i}\left(g^{jq}\partial_{j}g_{ip}\right)+\frac{1}{4}\partial_{i}\left(g_{jp}\tilde{\partial}^{j}g^{iq}\right)\nonumber \\
 &  & -\frac{1}{8}g_{ip}\tilde{\partial}^{i}g_{kl}\partial_{j}g^{kl}g^{jq}-\frac{1}{8}g_{ip}\tilde{\partial}^{i}g^{kl}\partial_{j}g_{kl}g^{jq}+\frac{1}{4}g_{ip}\partial_{j}g_{kl}\tilde{\partial}^{l}g^{ik}g^{jq}+\frac{1}{4}g_{ip}\tilde{\partial}^{i}g^{kl}\partial_{l}g_{jk}g^{jq}\nonumber \\
 &  & -\frac{1}{4}g_{ip}\tilde{\partial}^{l}\left(g_{kl}\partial_{j}g^{ik}\right)g^{jq}-\frac{1}{4}g_{ip}\partial_{l}\left(g^{kl}\tilde{\partial}^{i}g_{jk}\right)g^{jq}-\frac{1}{4}g_{ip}\tilde{\partial}^{l}\left(g^{ki}\partial_{k}g_{lj}\right)g^{jq}-\frac{1}{4}g_{ip}\partial_{l}\left(g_{kj}\tilde{\partial}^{k}g^{li}\right)g^{jq}\nonumber \\
 &  & 2\partial_{p}\tilde{\partial}^{q}d-\tilde{\partial}^{i}d\left(g^{jq}\partial_{j}g_{ip}\right)-\partial_{i}d\left(g_{jp}\tilde{\partial}^{j}g^{iq}\right)\nonumber \\
 &  & -\tilde{\partial}^{i}d\left(g_{ji}\partial_{p}g^{qj}\right)-\partial_{i}d\left(g^{ji}\tilde{\partial}^{q}g_{pj}\right)-2g_{ip}\tilde{\partial}^{i}\partial_{j}dg^{jq}\nonumber \\
 &  & +\frac{1}{2}g_{ip}\tilde{\partial}^{l}d\left(g_{kl}\partial_{j}g^{ik}\right)g^{jq}+\frac{1}{2}g_{ip}\partial_{l}d\left(g^{kl}\tilde{\partial}^{i}g_{jk}\right)g^{jq}\nonumber \\
 &  & +\frac{1}{2}g_{ip}\tilde{\partial}^{l}d\left(g^{ki}\partial_{k}g_{lj}\right)g^{jq}+\frac{1}{2}g_{ip}\partial_{l}d\left(g_{kj}\tilde{\partial}^{k}g^{li}\right)g^{jq}.\nonumber \\
\label{eq: GRA EOM}
\end{eqnarray}

\noindent On the other hand, the EOM for the dilaton is given by:

\begin{eqnarray}
 &  & \frac{1}{8}g_{ij}\tilde{\partial}^{i}g_{kl}\tilde{\partial}^{j}g^{kl}+\frac{1}{8}g_{ij}\tilde{\partial}^{i}g^{kl}\tilde{\partial}^{j}g_{kl}+\frac{1}{8}g^{ij}\partial_{i}g_{kl}\partial_{j}g^{kl}+\frac{1}{8}g^{ij}\partial_{i}g^{kl}\partial_{j}g_{kl}\nonumber \\
 &  & -\frac{1}{2}g_{ij}\tilde{\partial}^{i}g_{kl}\tilde{\partial}^{k}g^{jl}-\frac{1}{2}g^{ij}\partial_{i}g^{kl}\partial_{k}g_{jl}-\tilde{\partial}^{i}\tilde{\partial}^{j}g_{ij}-\partial_{i}\partial_{j}g^{ij}-4g_{ij}\tilde{\partial}^{i}d\tilde{\partial}^{j}d-4g^{ij}\partial_{i}d\partial_{j}d\nonumber \\
 &  & +4\tilde{\partial}^{i}g_{ij}\tilde{\partial}^{j}d+4\partial_{i}g^{ij}\partial_{j}d+4g_{ij}\tilde{\partial}^{i}\tilde{\partial}^{j}d+4g^{ij}\partial_{i}\partial_{j}d=0.\label{eq:DILA EOM}
\end{eqnarray}

\noindent Given the symmetries between the generalized Ricci scalars
$\mathcal{R}_{M\left(p\right)N\left(q\right)}$:

\begin{eqnarray}
\mathcal{R}_{\mathbf{1}\left(p\right)\mathbf{1}\left(q\right)} & \underleftrightarrow{g^{\bullet\bullet}\leftrightarrow g_{\bullet\bullet},\quad\tilde{\partial}^{\bullet}\leftrightarrow\partial_{\bullet}} & \mathcal{R}_{\mathbf{2}\left(p\right)\mathbf{2}\left(q\right)},\nonumber \\
\mathcal{R}_{\mathbf{1}\left(p\right)\mathbf{2}\left(q\right)} & \underleftrightarrow{g^{\bullet\bullet}\leftrightarrow g_{\bullet\bullet},\quad\tilde{\partial}^{\bullet}\leftrightarrow\partial_{\bullet}} & \mathcal{R}_{\mathbf{2}\left(p\right)\mathbf{1}\left(q\right)},
\end{eqnarray}

\noindent we need only compute one set. In this paper, we focus on
calculating $\mathcal{R}_{\mathbf{2}\left(p\right)\mathbf{2}\left(q\right)}$
and $\mathcal{R}_{\mathbf{2}\left(p\right)\mathbf{1}\left(q\right)}$.

Next, we present three types of black hole solutions for the EOM (\ref{eq: GRA EOM})
and (\ref{eq:DILA EOM}). To begin, we revisit the low-energy effective
action of closed string theory:

\begin{equation}
S=\int d^{4}x\sqrt{-g}e^{-2\phi}\left[R+4\left(\nabla\phi\right)^{2}-\frac{1}{12}H_{\mu\nu\rho}H^{\mu\nu\rho}\right].\label{eq:effective action}
\end{equation}

\noindent The solutions of standard DFT black holes are categorized
by the solutions $g_{ij}\left(x\right)$ and $\phi\left(x\right)$
from the action (\ref{eq:effective action}), as follows:

\vspace*{4.0ex}

\subsubsection*{1st kind of black hole solutions}

If $g_{ij}\left(x\right)$ and $\phi\left(x\right)$ satisfy the action
(\ref{eq:effective action}), they automatically become solutions
of DFT that satisfy the strong constraint. The corresponding line
elements can take two forms:

\begin{equation}
ds^{2}=g^{ij}\left(x\right)d\tilde{x}_{i}d\tilde{x}_{j}+g_{ij}\left(x\right)dx^{i}dx^{j},\qquad or\qquad ds^{2}=\left(g^{-1}\right)^{ij}\left(\tilde{x}\right)d\tilde{x}_{i}d\tilde{x}_{j}+\left(g^{-1}\right)_{ij}\left(\tilde{x}\right)dx^{i}dx^{j}.
\end{equation}

\noindent Moreover, when $\phi\left(x\right)=0$, the low-energy effective
action (\ref{eq:effective action}) reduces to the Einstein-Hilbert
action, implying that DFT also includes the black hole solutions of
Einstein's gravity. There is a notable exception: the BTZ black hole.
The BTZ black hole cannot be directly obtained by setting $\phi\left(x\right)=0$;
it also requires the inclusion of the Kalb-Ramond field ($H_{ijk}\neq0$)
to precisely reduce to Einstein\textquoteright s equations with a
negative cosmological constant \cite{Horowitz:1993jc}. Therefore,
it is necessary to consider a non-vanishing $b_{ij}$ in this case.

\noindent \textbf{Example:}

\noindent To illustrate, consider the following ansatz:

\begin{eqnarray}
dS^{2} & = & -A\left(\tilde{r},r\right)^{-1}d\tilde{t}^{2}+B\left(\tilde{r},r\right)^{-1}d\tilde{r}^{2}+C\left(\tilde{r},r\right)^{-1}\left(d\tilde{\theta}^{2}+D\left(\tilde{\theta},\theta\right)^{-1}d\tilde{\phi}^{2}\right)\nonumber \\
 &  & -A\left(\tilde{r},r\right)dt^{2}+B\left(\tilde{r},r\right)dr^{2}+C\left(\tilde{r},r\right)\left(d\theta^{2}+D\left(\tilde{\theta},\theta\right)d\phi^{2}\right).
\end{eqnarray}

\noindent Substituting this ansatz into the EOM (\ref{eq: GRA EOM})
and (\ref{eq:DILA EOM}), we obtain the EOM for the graviton:

\begin{eqnarray}
B\left[\left(\frac{\tilde{\partial}^{r}B}{B}-\frac{\tilde{\partial}^{r}A}{A}\right)\frac{\tilde{\partial}^{r}A}{A}+\frac{\tilde{\partial}^{r}\tilde{\partial}^{r}A}{A}-2\tilde{\partial}^{r}d\frac{\tilde{\partial}^{r}A}{A}\right]-\frac{1}{B}\left[\left(\frac{\partial_{r}B}{B}+\frac{\partial_{r}A}{A}\right)\frac{\partial_{r}A}{A}-\frac{\partial_{r}\partial_{r}A}{A}+2\partial_{r}d\frac{\partial_{r}A}{A}\right] & = & 0,\nonumber \\
\nonumber \\
B\left[\frac{1}{2}\left(\frac{\tilde{\partial}^{r}A}{A}\right)^{2}-\frac{1}{2}\left(\frac{\tilde{\partial}^{r}B}{B}\right)^{2}+\left(\frac{\tilde{\partial}^{r}C}{C}\right)^{2}+\frac{\tilde{\partial}^{r}\tilde{\partial}^{r}B}{B}-2\tilde{\partial}^{r}d\frac{\tilde{\partial}^{r}B}{B}-4\tilde{\partial}^{r}\tilde{\partial}^{r}d\right]\nonumber \\
-\frac{1}{B}\left[\frac{1}{2}\left(\frac{\partial_{r}A}{A}\right)^{2}+\frac{3}{2}\left(\frac{\partial_{r}B}{B}\right)^{2}+\left(\frac{\partial_{r}C}{C}\right)^{2}-\frac{\partial_{r}\partial_{r}B}{B}+2\partial_{r}d\frac{\partial_{r}B}{B}-4\partial_{r}\partial_{r}d\right] & = & 0,\nonumber \\
\nonumber \\
\frac{1}{2}B\left[-\frac{\tilde{\partial}^{r}B}{B}\frac{\tilde{\partial}^{r}C}{C}-\frac{\tilde{\partial}^{r}\tilde{\partial}^{r}C}{C}+\left(\frac{\tilde{\partial}^{r}C}{C}\right)^{2}+2\tilde{\partial}^{r}d\frac{\tilde{\partial}^{r}C}{C}\right]-C\left[2\tilde{\partial}^{\theta}\tilde{\partial}^{\theta}d-\frac{1}{4}\left(\frac{\tilde{\partial}^{\theta}D}{D}\right)^{2}\right]\nonumber \\
-\frac{1}{2}\frac{1}{B}\left[-\frac{\partial_{r}B}{B}\frac{\partial_{r}C}{C}+\frac{\partial_{r}\partial_{r}C}{C}-\left(\frac{\partial_{r}C}{C}\right)^{2}-2\partial_{r}d\frac{\partial_{r}C}{C}\right]-\frac{1}{C}\left[-2\partial_{\theta}\partial_{\theta}d+\frac{1}{4}\left(\frac{\partial_{\theta}D}{D}\right)^{2}\right] & = & 0,\nonumber \\
\nonumber \\
\frac{1}{2}B\left[-\frac{\tilde{\partial}^{r}B}{B}\frac{\tilde{\partial}^{r}C}{C}-\frac{\tilde{\partial}^{r}\tilde{\partial}^{r}C}{C}+\left(\frac{\tilde{\partial}^{r}C}{C}\right)^{2}+2\tilde{\partial}^{r}d\frac{\tilde{\partial}^{r}C}{C}\right]-C\left[\frac{1}{2}\frac{\tilde{\partial}^{\theta}\tilde{\partial}^{\theta}D}{D}-\frac{1}{2}\left(\frac{\tilde{\partial}^{\theta}D}{D}\right)^{2}-\tilde{\partial}^{\theta}d\frac{\tilde{\partial}^{\theta}D}{D}\right]\nonumber \\
-\frac{1}{2}\frac{1}{B}\left[-\frac{\partial_{r}B}{B}\frac{\partial_{r}C}{C}+\frac{\partial_{r}\partial_{r}C}{C}-\left(\frac{\partial_{r}C}{C}\right)^{2}-2\partial_{r}d\frac{\partial_{r}C}{C}\right]-\frac{1}{C}\left[\frac{1}{2}\frac{\partial_{\theta}\partial_{\theta}D}{D}-\frac{1}{2}\left(\frac{\partial_{\theta}D}{D}\right)^{2}-\partial_{\theta}d\frac{\partial_{\theta}D}{D}\right] & = & 0,\nonumber \\
\end{eqnarray}

\noindent and the EOM for the dilaton:

\begin{eqnarray}
 &  & \frac{1}{2}B\left(-\frac{1}{2}\left(\frac{\tilde{\partial}^{r}A}{A}\right)^{2}+\frac{1}{2}\left(\frac{\tilde{\partial}^{r}B}{B}\right)^{2}-\left(\frac{\tilde{\partial}^{r}C}{C}\right)^{2}-2\frac{\tilde{\partial}^{r}\tilde{\partial}^{r}B}{B}+8\frac{\tilde{\partial}^{r}B}{B}\tilde{\partial}^{r}d+8\tilde{\partial}^{r}\tilde{\partial}^{r}d-8\tilde{\partial}^{r}d\tilde{\partial}^{r}d\right)\nonumber \\
 &  & +C\left[4\tilde{\partial}^{\theta}\tilde{\partial}^{\theta}d-4\tilde{\partial}^{\theta}d\tilde{\partial}^{\theta}d-\frac{1}{4}\left(\frac{\tilde{\partial}^{\theta}D}{D}\right)^{2}\right]\nonumber \\
 &  & +\frac{1}{2}\frac{1}{B}\left(-\frac{1}{2}\left(\frac{\partial_{r}A}{A}\right)^{2}-\frac{7}{2}\left(\frac{\partial_{r}B}{B}\right)^{2}-\left(\frac{\partial_{r}C}{C}\right)^{2}+2\frac{\partial_{r}\partial_{r}B}{B}-8\frac{\partial_{r}B}{B}\partial_{r}d+8\partial_{r}\partial_{r}d-8\partial_{r}d\partial_{r}d\right)\nonumber \\
 &  & +\frac{1}{C}\left[4\partial_{\theta}\partial_{\theta}d-4\partial_{\theta}d\partial_{\theta}d-\frac{1}{4}\left(\frac{\partial_{\theta}D}{D}\right)^{2}\right]=0.
\end{eqnarray}

\noindent Keep in mind that $\tilde{\partial}^{r}\equiv\frac{\partial}{\partial\tilde{r}}$
and $\tilde{\partial}^{\theta}\equiv\frac{\partial}{\partial\tilde{\theta}}$.
After imposing the strong constraint, the solutions are organized
as follows. The first solution is:

\begin{eqnarray}
dS^{2} & = & -A\left(r\right)^{-1}d\tilde{t}^{2}+B\left(r\right)^{-1}d\tilde{r}^{2}+C\left(r\right)^{-1}\left(d\tilde{\theta}^{2}+D\left(\theta\right)^{-1}d\tilde{\phi}^{2}\right)\nonumber \\
 &  & -A\left(r\right)dt^{2}+B\left(r\right)dr^{2}+C\left(r\right)\left(d\theta^{2}+D\left(\theta\right)d\phi^{2}\right),
\end{eqnarray}

\noindent where

\begin{eqnarray}
A\left(r\right) & = & \left(1-\frac{2\eta}{r}\right)^{\frac{m+\sigma}{\eta}},\nonumber \\
B\left(r\right) & = & \left(1-\frac{2\eta}{r}\right)^{\frac{\sigma-m}{\eta}},\nonumber \\
C\left(r\right) & = & \left(1-\frac{2\eta}{r}\right)^{1+\frac{\sigma-m}{\eta}}r^{2},\nonumber \\
D\left(\theta\right) & = & \sin^{2}\theta,\nonumber \\
d\left(r,\theta\right) & = & \frac{\sigma}{2\eta}\ln\left(1-\frac{2\eta}{r}\right)-\frac{1}{4}\log\left[r^{4}\left(1-\frac{2\eta}{r}\right)^{2+3\frac{\sigma-m}{\eta}+\frac{m+\sigma}{\eta}}\sin^{2}\theta\right].
\end{eqnarray}

\noindent Its dual solution can be obtained similarly using the following
ansatz:

\begin{eqnarray}
dS^{2} & = & -A\left(\tilde{r}\right)d\tilde{t}^{2}+B\left(\tilde{r}\right)d\tilde{r}^{2}+C\left(\tilde{r}\right)\left(d\tilde{\theta}^{2}+D\left(\tilde{\theta}\right)d\tilde{\phi}^{2}\right)\nonumber \\
 &  & -A\left(\tilde{r}\right)^{-1}dt^{2}+B\left(\tilde{r}\right)^{-1}dr^{2}+C\left(\tilde{r}\right)^{-1}\left(d\theta^{2}+D\left(\tilde{\theta}\right)^{-1}d\phi^{2}\right),
\end{eqnarray}

\noindent and its solution is

\begin{eqnarray}
A\left(\tilde{r}\right) & = & \left(1-\frac{2\eta}{\tilde{r}}\right)^{\frac{m+\sigma}{\eta}},\nonumber \\
B\left(\tilde{r}\right) & = & \left(1-\frac{2\eta}{\tilde{r}}\right)^{\frac{\sigma-m}{\eta}},\nonumber \\
C\left(\tilde{r}\right) & = & \left(1-\frac{2\eta}{\tilde{r}}\right)^{1+\frac{\sigma-m}{\eta}}\tilde{r}^{2},\nonumber \\
D\left(\tilde{\theta}\right) & = & \sin^{2}\tilde{\theta},\nonumber \\
d\left(\tilde{r},\tilde{\theta}\right) & = & \frac{\sigma}{2\eta}\ln\left(1-\frac{2\eta}{\tilde{r}}\right)-\frac{1}{4}\log\left[\tilde{r}^{4}\left(1-\frac{2\eta}{\tilde{r}}\right)^{2+3\frac{\sigma-m}{\eta}+\frac{m+\sigma}{\eta}}\sin^{2}\tilde{\theta}\right],
\end{eqnarray}

\noindent where $m$ is the mass of the black hole, $\sigma$ is the
scalar charge, and $\eta^{2}=m^{2}+\sigma^{2}$. When $\sigma=0$,
these reduce to doubly Schwarzschild black holes. A similar solution
in the low-energy effective action was found in \cite{Kar:1998rv}.

\vspace*{4.0ex}

\subsubsection*{2nd kind of black hole solutions}

If $g_{ij}\left(x\right)$ and $\phi\left(x\right)$ are solutions
of the action (\ref{eq:effective action}), and $\left(g^{-1}\right)_{ij}\left(x\right)$
is also a solution for the same action. They satisfy the strong constraint
and encompass four distinct types, expressed as:

\begin{equation}
ds^{2}=g^{ij}\left(x\right)d\tilde{x}_{i}d\tilde{x}_{j}+g_{ij}\left(x\right)dx^{i}dx^{j},\qquad or\qquad ds^{2}=\left(g^{-1}\right)^{ij}\left(x\right)d\tilde{x}_{i}d\tilde{x}_{j}+\left(g^{-1}\right)_{ij}\left(x\right)dx^{i}dx^{j}{\color{brown},}
\end{equation}

\noindent or

\begin{equation}
ds^{2}=g^{ij}\left(\tilde{x}\right)d\tilde{x}_{i}d\tilde{x}_{j}+g_{ij}\left(\tilde{x}\right)dx^{i}dx^{j},\qquad or\qquad ds^{2}=\left(g^{-1}\right)^{ij}\left(\tilde{x}\right)d\tilde{x}_{i}d\tilde{x}_{j}+\left(g^{-1}\right)_{ij}\left(\tilde{x}\right)dx^{i}dx^{j}.
\end{equation}

\noindent \textbf{Example:}

\noindent To illustrate this property, consider the following ansatz:

\begin{eqnarray}
dS^{2} & = & -A\left(\tilde{r},r\right)^{-1}d\tilde{t}^{2}+d\tilde{r}^{2}+A\left(\tilde{r},r\right)^{-1}\left(d\tilde{x}_{i}d\tilde{x}^{i}\right)\nonumber \\
 &  & -A\left(\tilde{r},r\right)dt^{2}+dr^{2}+A\left(\tilde{r},r\right)\left(dx_{i}dx^{i}\right).
\end{eqnarray}

\noindent The EOM (\ref{eq: GRA EOM}) and (\ref{eq:DILA EOM}) for
the graviton are then:

\begin{eqnarray}
\left[-\left(\frac{\tilde{\partial}^{r}A}{A}\right)^{2}+\frac{\tilde{\partial}^{r}\tilde{\partial}^{r}A}{A}-2\tilde{\partial}^{r}d\frac{\tilde{\partial}^{r}A}{A}\right]-\left[\left(\frac{\partial_{r}A}{A}\right)^{2}-\frac{\partial_{r}\partial_{r}A}{A}+2\partial_{r}d\frac{\partial_{r}A}{A}\right] & = & 0,\nonumber \\
\nonumber \\
\left[\frac{D-1}{2}\left(\frac{\tilde{\partial}^{r}A}{A}\right)^{2}-4\tilde{\partial}^{r}\tilde{\partial}^{r}d\right]-\left[\frac{D-1}{2}\left(\frac{\partial_{r}A}{A}\right)^{2}-4\partial_{r}\partial_{r}d\right] & = & 0,
\end{eqnarray}

\noindent while the EOM for the dilaton is:

\begin{equation}
\left(-\frac{D-1}{2}\left(\frac{\tilde{\partial}^{r}A}{A}\right)^{2}+8\tilde{\partial}^{r}\tilde{\partial}^{r}d-8\tilde{\partial}^{r}d\tilde{\partial}^{r}d\right)+\left(-\frac{D-1}{2}\left(\frac{\partial_{r}A}{A}\right)^{2}+8\partial_{r}\partial_{r}d-8\partial_{r}d\partial_{r}d\right)=0.
\end{equation}

\noindent This formulation yields two types of solutions, given by:

\begin{equation}
dS^{2}=-\left(\frac{r}{r_{0}}\right)^{\frac{\pm2}{\sqrt{D-1}}}d\tilde{t}^{2}+d\tilde{r}^{2}+\left(\frac{r}{r_{0}}\right)^{\frac{\pm2}{\sqrt{D-1}}}\left(d\tilde{x}_{i}d\tilde{x}^{i}\right)-\left(\frac{r}{r_{0}}\right)^{\frac{\mp2}{\sqrt{D-1}}}dt^{2}+dr^{2}+\left(\frac{r}{r_{0}}\right)^{\frac{\mp2}{\sqrt{D-1}}}\left(dx_{i}dx^{i}\right),
\end{equation}

\noindent which depend on the usual coordinate $r$, and two additional
types of solutions that depend on the dual coordinate $\tilde{r}$:

\begin{equation}
dS^{2}=-\left(\frac{\tilde{r}}{\tilde{r_{0}}}\right)^{\frac{\pm2}{\sqrt{D-1}}}d\tilde{t}^{2}+d\tilde{r}^{2}+\left(\frac{\tilde{r}}{\tilde{r_{0}}}\right)^{\frac{\pm2}{\sqrt{D-1}}}\left(d\tilde{x}_{i}d\tilde{x}^{i}\right)-\left(\frac{\tilde{r}}{\tilde{r_{0}}}\right)^{\frac{\mp2}{\sqrt{D-1}}}dt^{2}+dr^{2}+\left(\frac{\tilde{r}}{\tilde{r_{0}}}\right)^{\frac{\mp2}{\sqrt{D-1}}}\left(dx_{i}dx^{i}\right).
\end{equation}

\noindent In cosmology, such solutions exhibit scale-factor duality.

\vspace*{4.0ex}

\subsubsection*{3rd kind of black hole solutions}

This kind of solution in DFT involves double coordinates that are
treated simultaneously, leading to a violation of the strong constraint.
These solutions can be constructed using the product of self-dual
solutions: $g\left(\tilde{x},x\right)=g\left(\tilde{x}\right)\times\left(g^{-1}\right)\left(x\right)$.
The line element is expressed as:

\begin{equation}
ds^{2}=g^{ij}\left(x,\tilde{x}\right)d\tilde{x}_{i}d\tilde{x}_{j}+g_{ij}\left(x,\tilde{x}\right)dx^{i}dx^{j},\qquad or\qquad ds^{2}=\left(g^{-1}\right)^{ij}\left(x,\tilde{x}\right)d\tilde{x}_{i}d\tilde{x}_{j}+\left(g^{-1}\right)_{ij}\left(x,\tilde{x}\right)dx^{i}dx^{j}
\end{equation}

\noindent \textbf{Example:}

\noindent Consider the following ansatz:

\begin{eqnarray}
dS^{2} & = & -A\left(\tilde{r},r\right)^{-1}d\tilde{t}^{2}+B\left(\tilde{r},r\right)^{-1}d\tilde{r}^{2}+C\left(\tilde{r},r\right)^{-1}\left(d\tilde{x}_{i}d\tilde{x}^{i}\right)-A\left(\tilde{r},r\right)dt^{2}+B\left(\tilde{r},r\right)dr^{2}+C\left(\tilde{r},r\right)\left(dx_{i}dx^{i}\right).\nonumber \\
\end{eqnarray}

\noindent Using this ansatz, the corresponding EOM for the graviton
(\ref{eq: GRA EOM}) are:

\begin{eqnarray}
B\left[\left(\frac{\tilde{\partial}^{r}B}{B}-\frac{\tilde{\partial}^{r}A}{A}\right)\frac{\tilde{\partial}^{r}A}{A}+\frac{\tilde{\partial}^{r}\tilde{\partial}^{r}A}{A}-2\tilde{\partial}^{r}d\frac{\tilde{\partial}^{r}A}{A}\right]-\frac{1}{B}\left[\left(\frac{\partial_{r}B}{B}+\frac{\partial_{r}A}{A}\right)\frac{\partial_{r}A}{A}-\frac{\partial_{r}\partial_{r}A}{A}+2\partial_{r}d\frac{\partial_{r}A}{A}\right] & = & 0,\nonumber \\
\nonumber \\
B\left[\frac{1}{2}\left(\frac{\tilde{\partial}^{r}A}{A}\right)^{2}-\frac{1}{2}\left(\frac{\tilde{\partial}^{r}B}{B}\right)^{2}+\frac{D-2}{2}\left(\frac{\tilde{\partial}^{r}C}{C}\right)^{2}+\frac{\tilde{\partial}^{r}\tilde{\partial}^{r}B}{B}-2\tilde{\partial}^{r}d\frac{\tilde{\partial}^{r}B}{B}-4\tilde{\partial}^{r}\tilde{\partial}^{r}d\right]\nonumber \\
-\frac{1}{B}\left[\frac{1}{2}\left(\frac{\partial_{r}A}{A}\right)^{2}+\frac{3}{2}\left(\frac{\partial_{r}B}{B}\right)^{2}+\frac{D-2}{2}\left(\frac{\partial_{r}C}{C}\right)^{2}-\frac{\partial_{r}\partial_{r}B}{B}+2\partial_{r}d\frac{\partial_{r}B}{B}-4\partial_{r}\partial_{r}d\right] & = & 0,\nonumber \\
\nonumber \\
B\left[-\frac{\tilde{\partial}^{r}B}{B}\frac{\tilde{\partial}^{r}C}{C}-\frac{\tilde{\partial}^{r}\tilde{\partial}^{r}C}{C}+\left(\frac{\tilde{\partial}^{r}C}{C}\right)^{2}+2\tilde{\partial}^{r}d\frac{\tilde{\partial}^{r}C}{C}\right]-\frac{1}{B}\left[-\frac{\partial_{r}B}{B}\frac{\partial_{r}C}{C}+\frac{\partial_{r}\partial_{r}C}{C}-\left(\frac{\partial_{r}C}{C}\right)^{2}-2\partial_{r}d\frac{\partial_{r}C}{C}\right] & = & 0,\nonumber \\
\end{eqnarray}

\noindent Similarily, we derive the EOM for the dilaton (\ref{eq:DILA EOM}):

\begin{eqnarray}
B\left(-\frac{1}{2}\left(\frac{\tilde{\partial}^{r}A}{A}\right)^{2}+\frac{1}{2}\left(\frac{\tilde{\partial}^{r}B}{B}\right)^{2}-\frac{D-2}{2}\left(\frac{\tilde{\partial}^{r}C}{C}\right)^{2}-2\frac{\tilde{\partial}^{r}\tilde{\partial}^{r}B}{B}+8\frac{\tilde{\partial}^{r}B}{B}\tilde{\partial}^{r}d+8\tilde{\partial}^{r}\tilde{\partial}^{r}d-8\tilde{\partial}^{r}d\tilde{\partial}^{r}d\right)\nonumber \\
+\frac{1}{B}\left(-\frac{1}{2}\left(\frac{\partial_{r}A}{A}\right)^{2}-\frac{7}{2}\left(\frac{\partial_{r}B}{B}\right)^{2}-\frac{D-2}{2}\left(\frac{\partial_{r}C}{C}\right)^{2}+2\frac{\partial_{r}\partial_{r}B}{B}-8\frac{\partial_{r}B}{B}\partial_{r}d+8\partial_{r}\partial_{r}d-8\partial_{r}d\partial_{r}d\right) & = & 0.\nonumber \\
\end{eqnarray}

\noindent The solutions are expressed as:

\begin{eqnarray}
dS^{2} & = & -A\left(\tilde{r},r\right)^{\mp1}d\tilde{t}^{2}+B\left(\tilde{r},r\right)^{\mp1}d\tilde{r}^{2}+C\left(\tilde{r},r\right)^{\mp1}\left(d\tilde{x}_{i}d\tilde{x}^{i}\right)\nonumber \\
 &  & -A\left(\tilde{r},r\right)^{\pm}dt^{2}+B\left(\tilde{r},r\right)^{\pm}dr^{2}+C\left(\tilde{r},r\right)^{\pm}\left(dx_{i}dx^{i}\right),
\end{eqnarray}

\noindent where

\begin{eqnarray}
A\left(\tilde{r},r\right) & = & \left(r+a\right)^{\frac{1}{\sqrt{3}}}\left(\tilde{r}+a\right)^{-\frac{1}{\sqrt{3}}},\nonumber \\
B\left(\tilde{r},r\right) & = & \left(r+a\right)^{-1}\left(\tilde{r}+a\right),\nonumber \\
C\left(\tilde{r},r\right) & = & \left(r+a\right)^{\sqrt{\frac{2}{3\left(D-2\right)}}}\left(\tilde{r}+a\right)^{-\sqrt{\frac{2}{3\left(D-2\right)}}},\nonumber \\
d\left(\tilde{r},\tilde{\theta},r,\theta\right) & = & const.
\end{eqnarray}

\section{Space/time split double field theory}

In the standard DFT, all spacetime coordinates are doubled from $D$
to $2D$, resulting in the presence of double time coordinates. This
feature complicates the physical interpretation of DFT solutions.
A natural resolution to this issue is to undouble the time coordinate,
meaning that all fields in DFT depend on a single time coordinate
while the spatial coordinates are doubled. The action for this $1+2\left(D-1\right)$-dimensional
spacetime can be derived through an ADM-like decomposition of the
generalized metric and the dilaton \cite{Hohm:2013nja,Naseer:2015fba,Hohm:2022pfi}.
To perform this decomposition, we introduce the following notation:
the double coordinates $X^{M}$ can be rewritten as

\begin{equation}
X^{M}=\left(\tilde{t},t,X^{\hat{M}}\right),
\end{equation}

\noindent where the hat indices '$\hat{M}$' denote the spatial part
of the $2D$ manifold and the tilde indices '$\tilde{t}$' represent
the dual coordinates. Similarly, the generalized metric $\mathcal{H}^{MN}$
and the invariant metric $\eta^{MN}$ can be decomposed as

\begin{equation}
\mathcal{H}^{MN}=\left(\begin{array}{ccc}
\mathcal{H}^{\tilde{t}\tilde{t}} & 0 & 0\\
0 & \mathcal{H}^{tt} & 0\\
0 & 0 & \mathcal{H}^{\hat{M}\hat{N}}
\end{array}\right),\qquad\eta^{MN}=\left(\begin{array}{ccc}
0 & 1 & 0\\
1 & 0 & 0\\
0 & 0 & \eta^{\hat{M}\hat{N}}
\end{array}\right),\label{eq:metric decomposition}
\end{equation}

\noindent where we set the generalized shift vector to zero, thus
$\mathcal{H}^{tM}$=$\mathcal{H}^{\tilde{t}M}$=0, $\mathcal{H}^{tt}=-n^{-2}$,
$\mathcal{H}^{\tilde{t}\tilde{t}}=-n^{2}$ and $n$ is the lapse function.
The dilaton can then be rewritten as

\begin{equation}
e^{-2d}=ne^{-2\hat{d}}.\label{eq:dilaton decomposition}
\end{equation}

\noindent Substituting the decompositions (\ref{eq:metric decomposition})
and (\ref{eq:dilaton decomposition}) back into the DFT action, we
obtain:

\begin{equation}
S_{\mathrm{DFT}}=\int d^{D}xd^{D}\tilde{x}\mathcal{L}_{\mathrm{DFT}}\left(d,\mathcal{H}_{MN}\right),
\end{equation}

\noindent where

\begin{eqnarray}
\mathcal{L}_{\mathrm{DFT}}\left(d,\mathcal{H}_{MN}\right) & = & e^{-2d}\left(\frac{1}{8}\mathcal{H}^{MN}\partial_{M}\mathcal{H}^{KL}\partial_{N}\mathcal{H}_{KL}-\frac{1}{2}\mathcal{H}^{MN}\partial_{N}\mathcal{H}^{KL}\partial_{L}\mathcal{H}_{MK}\right.\nonumber \\
 &  & \left.-2\partial_{M}d\partial_{N}\mathcal{H}^{MN}+4\mathcal{H}^{MN}\partial_{M}d\partial_{N}d\right).\label{eq: DFT lagrangian}
\end{eqnarray}

\noindent By removing the dual time-like coordinate through the constraint
$\frac{\partial}{\partial\tilde{t}}\left(\cdots\right)=0$, we obtain:

\begin{equation}
S_{\mathrm{DFT}}=\int dt\int d^{D-1}xd^{D-1}\tilde{x}\mathcal{\hat{L}}_{\mathrm{DFT}},
\end{equation}

\noindent where

\begin{eqnarray}
\mathcal{\hat{L}}_{\mathrm{DFT}} & = & \mathcal{L}_{\mathrm{DFT}}\left(\hat{d}-\frac{1}{2}\log n,\mathcal{H}_{\hat{M}\hat{N}}\right)-\frac{1}{n}e^{-2\hat{d}}\left(4\left(\partial_{t}\hat{d}\right)^{2}+\frac{1}{8}\partial_{t}\mathcal{H}^{\hat{M}\hat{N}}\partial_{t}\mathcal{H}_{\hat{M}\hat{N}}+\mathcal{H}^{\hat{M}\hat{N}}\partial_{\hat{M}}n\partial_{\hat{N}}n\right).\nonumber \\
\end{eqnarray}

\noindent This implies that we replace $d$ as $\hat{d}-\frac{1}{2}\log n$
and $\mathcal{H}_{MN}$ with $\mathcal{H}_{\hat{M}\hat{N}}$ in the
Lagrangian (\ref{eq: DFT lagrangian}). This action is consistent
with \cite{Naseer:2015fba}. After a brief calculation, it also shows
that

\begin{equation}
\mathcal{L}_{\mathrm{DFT}}\left(\hat{d}-\frac{1}{2}\log n,\mathcal{H}_{\hat{M}\hat{N}}\right)-n^{-1}e^{-2d}\mathcal{H}^{\hat{M}\hat{N}}\partial_{\hat{M}}n\partial_{\hat{N}}n=ne^{-2d}\mathcal{R}\left(\hat{d},\mathcal{H}_{\hat{M}\hat{N}}\right)-b_{1},
\end{equation}

\noindent where $b_{1}$ is a total derivative term (boundary term):

\begin{equation}
b_{1}=-\partial_{\hat{M}}\left(ne^{-2\hat{d}}\left[\partial_{\hat{N}}\mathcal{H}^{\hat{M}\hat{N}}-4\mathcal{H}^{\hat{M}\hat{N}}\partial_{\hat{N}}\hat{d}\right]\right).
\end{equation}
Finally, the $1+2\left(D-1\right)$-dimensional DFT action is given
by

\begin{eqnarray}
S_{\mathrm{DFT}} & = & \int dtd^{D-1}xd^{D-1}\tilde{x}\mathcal{\hat{L}}_{\mathrm{DFT}}\nonumber \\
 & = & -\int dtd^{D-1}xd^{D-1}\tilde{x}\left[\frac{1}{n}e^{-2\hat{d}}\left(4\left(\partial_{t}\hat{d}\right)^{2}+\frac{1}{8}\partial_{t}\mathcal{H}^{\hat{M}\hat{N}}\partial_{t}\mathcal{H}_{\hat{M}\hat{N}}\right)-ne^{-2d}\mathcal{R}\left(\hat{d},\mathcal{H}_{\hat{M}\hat{N}}\right)\right].\label{eq:2d+1 action}
\end{eqnarray}

\noindent Since all decomposed fields of DFT only depend on $\left(t,X^{\hat{M}}\right)$:

\begin{equation}
\mathcal{H}^{\hat{M}\hat{N}}\left(t,X^{\hat{M}}\right),\qquad\hat{d}\left(t,X^{\hat{M}}\right),\qquad n\left(t,X^{\hat{M}}\right).
\end{equation}

\noindent and the spacetime action is an integral over a $1+2\left(D-1\right)$-dimensional
spacetime, the dual timelike direction $\tilde{t}$ has no physical
meaning. This implies that we can safely ignore the dual timelike
direction $\tilde{t}$, and the action (\ref{eq:2d+1 action}) can
be effectively considered as the action for a $1+2\left(D-1\right)$
spacetime with the following ansatz:

\begin{equation}
ds^{2}=-n^{2}dt^{2}+\mathcal{H}_{\hat{M}\hat{N}}dX^{\hat{M}}dX^{\hat{N}},
\end{equation}

\noindent which can be expanded as

\begin{eqnarray}
ds^{2} & = & -n^{2}dt^{2}+\mathcal{H}_{\hat{\mathbf{1}}\hat{\mathbf{1}}}dX^{\hat{\mathbf{1}}}dX^{\hat{\mathbf{1}}}+\mathcal{H}_{\hat{\mathbf{1}}\hat{\mathbf{2}}}dX^{\hat{\mathbf{1}}}dX^{\hat{\mathbf{2}}}+\mathcal{H}_{\hat{\mathbf{2}}\hat{\mathbf{1}}}dX^{\hat{\mathbf{2}}}dX^{\hat{\mathbf{1}}}+\mathcal{H}_{\hat{\mathbf{2}}\hat{\mathbf{2}}}dX^{\hat{\mathbf{2}}}dX^{\hat{\mathbf{2}}}\nonumber \\
 & = & -n^{2}dt^{2}+g^{ij}d\tilde{x}_{i}d\tilde{x}_{j}-g^{ik}b_{kj}d\tilde{x}_{i}dx^{j}+b_{ik}g^{kj}dx^{i}d\tilde{x}_{j}+\left(g_{ij}-b_{ik}g^{kl}b_{lj}\right)dx^{i}dx^{j}.
\end{eqnarray}

\noindent For simplicity in the rest of the paper, we set the Kalb-Ramond
field to zero, so the line element becomes:

\begin{equation}
ds^{2}=-n^{2}dt^{2}+\mathcal{H}_{\hat{\mathbf{1}}\hat{\mathbf{1}}}dX^{\hat{\mathbf{1}}}dX^{\hat{\mathbf{1}}}+\mathcal{H}_{\hat{\mathbf{2}}\hat{\mathbf{2}}}dX^{\hat{\mathbf{2}}}dX^{\hat{\mathbf{2}}}=-n^{2}dt^{2}+g^{ij}d\tilde{x}_{i}d\tilde{x}_{j}+g_{ij}dx^{i}dx^{j}.
\end{equation}

\noindent Thus, our aim is to calculate the black hole solutions for
this ansatz in the rest of the paper.

To obtain the EOM, we can also substitute the field decompositions
(\ref{eq:metric decomposition}) and (\ref{eq:dilaton decomposition})
into the equations (\ref{eq:EOM}). The dilatonic EOM (\ref{eq:DILA EOM})
then becomes:

\begin{eqnarray}
\frac{1}{8}\mathcal{H}^{\hat{M}\hat{N}}\partial_{\hat{M}}\mathcal{H}^{\hat{K}\hat{L}}\partial_{\hat{N}}\mathcal{H}_{\hat{K}\hat{L}}-\frac{1}{2}\mathcal{H}^{\hat{M}\hat{N}}\partial_{\hat{M}}\mathcal{H}^{\hat{K}\hat{L}}\partial_{\hat{K}}\mathcal{H}_{\hat{N}\hat{L}}-\partial_{\hat{M}}\partial_{\hat{N}}\mathcal{H}^{\hat{M}\hat{N}}\nonumber \\
-4\mathcal{H}^{\hat{M}\hat{N}}\partial_{\hat{M}}d\partial_{\hat{N}}d+4\partial_{\hat{M}}\mathcal{H}^{\hat{M}\hat{N}}\partial_{\hat{N}}d+4\mathcal{H}^{\hat{M}\hat{N}}\partial_{\hat{M}}\partial_{\hat{N}}d\nonumber \\
+\frac{1}{4}\frac{1}{n^{2}}\partial_{t}n^{-2}\partial_{t}n^{2}+\partial_{t}\partial_{t}n^{-2}-4\partial_{t}n^{-2}\partial_{t}d-\frac{1}{8}\frac{1}{n^{2}}\partial_{t}\mathcal{H}^{\hat{K}\hat{L}}\partial_{t}\mathcal{H}_{\hat{K}\hat{L}}+\frac{1}{4}\mathcal{H}^{\hat{M}\hat{N}}\partial_{\hat{M}}n^{-2}\partial_{\hat{N}}n^{2} & = & 0.\label{eq:DLA EOM 2}
\end{eqnarray}

\noindent The gravitational EOM (\ref{eq: GRA EOM}) are given by:

\begin{eqnarray}
\mathcal{R}_{tt}=\mathcal{K}_{tt}-S_{\quad t}^{P}\mathcal{K}_{PQ}S_{\quad t}^{Q} & = & 0,\nonumber \\
\mathcal{R}_{\hat{M}\hat{N}}=\mathcal{K}_{\hat{M}\hat{N}}-S_{\quad\hat{M}}^{P}\mathcal{K}_{PQ}S_{\quad\hat{N}}^{Q} & = & 0.\label{eq:GR EOM 2}
\end{eqnarray}

\noindent For simplicity, we will continue to use $d$ instead of
$\hat{d}$ in the rest of the paper.

\subsection{Black hole solutions}

To calculate the black hole solutions of space/time split DFT, we
utilize the following ansatz:

\begin{equation}
ds^{2}=-n^{2}dt^{2}+\mathcal{H}_{\hat{M}\hat{N}}dX^{\hat{M}}dX^{\hat{N}},
\end{equation}

\noindent where we assume the generalized metric with the vanishing
Kalb-Ramond field takes the form:

\begin{equation}
\mathcal{H}_{\hat{M}\left(i\right)\hat{N}\left(j\right)}\equiv\left(\begin{array}{cc}
\mathcal{H}_{\hat{\mathbf{1}}\left(i\right)\hat{\mathbf{1}}\left(j\right)} & 0\\
0 & \mathcal{H}_{\hat{\mathbf{2}}\left(i\right)\hat{\mathbf{2}}\left(j\right)}
\end{array}\right)=\left(\begin{array}{cc}
g^{ij} & 0\\
0 & g_{ij}
\end{array}\right),
\end{equation}

\noindent where $g_{ij}$ denotes the spatial section of spacetime.
Substituting this ansatz into the gravitational EOM (\ref{eq:GR EOM 2}),
the non-vanishing generalized Ricci scalar can be given by $\mathcal{R}_{\hat{M}\left(i\right)\hat{N}\left(j\right)}$
and $\mathcal{R}_{tt}$ which are given as follows:

\begin{equation}
\mathcal{R}_{\hat{M}\left(i\right)\hat{N}\left(j\right)}=\mathcal{K}_{\hat{M}\hat{N}}-S_{\quad\hat{M}}^{P}\mathcal{K}_{PQ}S_{\quad\hat{N}}^{Q}\equiv\left(\begin{array}{cc}
\mathcal{R}_{\hat{\mathbf{1}}\left(i\right)\hat{\mathbf{1}}\left(j\right)} & \mathcal{R}_{\hat{\mathbf{1}}\left(i\right)\hat{\mathbf{2}}\left(j\right)}\\
\mathcal{R}_{\hat{\mathbf{2}}\left(i\right)\hat{\mathbf{1}}\left(j\right)} & \mathcal{R}_{\hat{\mathbf{2}}\left(i\right)\hat{\mathbf{2}}\left(j\right)}
\end{array}\right),
\end{equation}

\noindent and

\begin{eqnarray}
\mathcal{R}_{tt} & = & \frac{1}{4}\partial_{t}g_{ij}\partial_{t}g^{ij}-\frac{1}{4}\partial_{t}n^{2}\partial_{t}n^{-2}+\frac{1}{4}\partial_{t}\left(n^{-2}\partial_{t}n^{2}\right)\nonumber \\
 &  & -\frac{1}{4}n^{4}\partial_{t}\left(n^{-2}\partial_{t}n^{-2}\right)-\frac{1}{2}\partial_{t}d\left(n^{-2}\partial_{t}n^{2}\right)+2\partial_{t}\partial_{t}d\nonumber \\
 &  & +\frac{1}{4}\tilde{\partial}^{i}\left(g_{ij}\tilde{\partial}^{j}n^{2}\right)-\frac{1}{4}n^{4}\tilde{\partial}^{i}\left(g_{ij}\tilde{\partial}^{j}n^{-2}\right)-\frac{1}{2}\tilde{\partial}^{i}d\left(g_{ij}\tilde{\partial}^{j}n^{2}\right)+\frac{1}{2}n^{4}\tilde{\partial}^{i}d\left(g_{ij}\tilde{\partial}^{j}n^{-2}\right)\nonumber \\
 &  & +\frac{1}{4}\partial_{i}\left(g^{ij}\partial_{j}n^{2}\right)-\frac{1}{4}n^{4}\partial_{i}\left(g^{ij}\partial_{j}n^{-2}\right)-\frac{1}{2}\partial_{i}d\left(g^{ij}\partial_{j}n^{2}\right)+\frac{1}{2}n^{4}\partial_{i}d\left(g^{ij}\partial_{j}n^{-2}\right),
\end{eqnarray}

\noindent Then, $\mathcal{R}_{\hat{M}\left(i\right)\hat{N}\left(j\right)}$
can be expanded as:

\begin{eqnarray}
\mathcal{R}_{\hat{\mathbf{1}}\left(p\right)\hat{\mathbf{1}}\left(q\right)} & = & \frac{1}{4}\tilde{\partial}^{p}n^{-2}\tilde{\partial}^{q}n^{2}-\frac{1}{4}g^{ip}\partial_{i}n^{2}\partial_{j}n^{-2}g^{jq}+\frac{1}{4}\partial_{t}\left(n^{-2}\partial_{t}g^{pq}\right)+\frac{1}{4}g^{ip}\partial_{t}\left(n^{-2}\partial_{t}n^{2}\right)g^{jq}\nonumber \\
 &  & -\frac{1}{2}\partial_{t}d\left(n^{-2}\partial_{t}g^{pq}\right)+\frac{1}{2}g^{ip}\partial_{t}d\left(n^{-2}\partial_{t}g_{ij}\right)g^{jq}\nonumber \\
 &  & +\frac{1}{4}\tilde{\partial}^{p}g^{ij}\tilde{\partial}^{q}g_{ij}-\frac{1}{4}\tilde{\partial}^{i}\left(g_{ij}\tilde{\partial}^{j}g^{pq}\right)-\frac{1}{4}\partial_{i}\left(g^{ij}\partial_{j}g^{pq}\right)\nonumber \\
 &  & -\frac{1}{4}\tilde{\partial}^{q}g_{ij}\tilde{\partial}^{j}g^{pi}-\frac{1}{4}\tilde{\partial}^{p}g_{ij}\tilde{\partial}^{j}g^{qi}+\frac{1}{4}\tilde{\partial}^{i}\left(g_{ji}\tilde{\partial}^{q}g^{pj}\right)+\frac{1}{4}\tilde{\partial}^{i}\left(g_{ji}\tilde{\partial}^{p}g^{qj}\right)\nonumber \\
 &  & +\frac{1}{4}\partial_{i}\left(g^{jp}\partial_{j}g^{iq}\right)+\frac{1}{4}\partial_{i}\left(g^{jq}\partial_{j}g^{ip}\right)-\frac{1}{8}g^{ip}\partial_{i}g_{kl}\partial_{j}g^{kl}g^{jq}-\frac{1}{8}g^{ip}\partial_{i}g^{kl}\partial_{j}g_{kl}g^{jq}\nonumber \\
 &  & +\frac{1}{4}g^{ip}\tilde{\partial}^{l}\left(g_{lk}\tilde{\partial}^{k}g_{ij}\right)g^{jq}+\frac{1}{4}g^{ip}\partial_{l}\left(g^{lk}\partial_{k}g_{ij}\right)g^{jq}+\frac{1}{4}g^{ip}\partial_{j}g^{kl}\partial_{l}g_{ik}g^{jq}+\frac{1}{4}g^{ip}\partial_{i}g^{kl}\partial_{l}g_{jk}g^{jq}\nonumber \\
 &  & -\frac{1}{4}g^{ip}\partial_{l}\left(g^{kl}\partial_{j}g_{ik}\right)g^{jq}-\frac{1}{4}g^{ip}\partial_{l}\left(g^{kl}\partial_{i}g_{jk}\right)g^{jq}-\frac{1}{4}g^{ip}\tilde{\partial}^{l}\left(g_{ki}\tilde{\partial}^{k}g_{lj}\right)g^{jq}-\frac{1}{4}g^{ip}\tilde{\partial}^{l}\left(g_{kj}\tilde{\partial}^{k}g_{li}\right)g^{jq}\nonumber \\
 &  & +\frac{1}{2}\tilde{\partial}^{i}d\left(g_{ij}\tilde{\partial}^{j}g^{pq}\right)+\frac{1}{2}\partial_{i}d\left(g^{ij}\partial_{j}g^{pq}\right)+2\tilde{\partial}^{p}\tilde{\partial}^{q}d\nonumber \\
 &  & -\frac{1}{2}\partial_{i}d\left(g^{jp}\partial_{j}g^{iq}\right)-\frac{1}{2}\partial_{i}d\left(g^{jq}\partial_{j}g^{ip}\right)-\frac{1}{2}\tilde{\partial}^{i}d\left(g_{ji}\tilde{\partial}^{q}g^{pj}\right)-\frac{1}{2}\tilde{\partial}^{i}d\left(g_{ji}\tilde{\partial}^{p}g^{qj}\right)\nonumber \\
 &  & -\frac{1}{2}g^{ip}\tilde{\partial}^{l}d\left(g_{lk}\tilde{\partial}^{k}g_{ij}\right)g^{jq}-\frac{1}{2}g^{ip}\partial_{l}d\left(g^{lk}\partial_{k}g_{ij}\right)g^{jq}-2g^{ip}\partial_{i}\partial_{j}dg^{jq}\nonumber \\
 &  & +\frac{1}{2}g^{ip}\partial_{l}d\left(g^{kl}\partial_{j}g_{ik}\right)g^{jq}+\frac{1}{2}g^{ip}\partial_{l}d\left(g^{kl}\partial_{i}g_{jk}\right)g^{jq}\nonumber \\
 &  & +\frac{1}{2}g^{ip}\tilde{\partial}^{l}d\left(g_{ki}\tilde{\partial}^{k}g_{lj}\right)g^{jq}+\frac{1}{2}g^{ip}\tilde{\partial}^{l}d\left(g_{kj}\tilde{\partial}^{k}g_{li}\right)g^{jq},\\
\nonumber \\
\mathcal{R}_{\hat{\mathbf{2}}\left(p\right)\hat{\mathbf{2}}\left(q\right)} & = & \frac{1}{4}\partial_{p}n^{2}\partial_{q}n^{-2}-\frac{1}{4}g_{pp}\tilde{\partial}^{p}n^{2}\tilde{\partial}^{q}n^{-2}g_{qq}+\frac{1}{4}\partial_{t}\left(n^{-2}\partial_{t}g_{pq}\right)+\frac{1}{4}g_{ip}\partial_{t}\left(n^{-2}\partial_{t}g^{ij}\right)g_{jq}\nonumber \\
 &  & -\frac{1}{2}g_{ip}\partial_{t}d\left(n^{-2}\partial_{t}g^{ij}\right)g_{jq}+\frac{1}{2}\partial_{t}d\left(n^{-2}\partial_{t}g_{pq}\right)\nonumber \\
 &  & +\frac{1}{4}\partial_{p}g^{ij}\partial_{q}g_{ij}-\frac{1}{4}\tilde{\partial}^{i}\left(g_{ij}\tilde{\partial}^{j}g_{pq}\right)-\frac{1}{4}\partial_{i}\left(g^{ij}\partial_{j}g_{pq}\right)\nonumber \\
 &  & -\frac{1}{4}\partial_{q}g^{ij}\partial_{j}g_{pi}-\frac{1}{4}\partial_{p}g^{ij}\partial_{j}g_{qi}+\frac{1}{4}\partial_{i}\left(g^{ji}\partial_{q}g_{pj}\right)+\frac{1}{4}\partial_{i}\left(g^{ji}\partial_{p}g_{qj}\right)\nonumber \\
 &  & +\frac{1}{4}\tilde{\partial}^{i}\left(g_{jp}\tilde{\partial}^{j}g_{iq}\right)+\frac{1}{4}\tilde{\partial}^{i}\left(g_{jq}\tilde{\partial}^{j}g_{ip}\right)-\frac{1}{8}g_{ip}\tilde{\partial}^{i}g_{kl}\tilde{\partial}^{j}g^{kl}g_{jq}-\frac{1}{8}g_{ip}\tilde{\partial}^{i}g^{kl}\tilde{\partial}^{j}g_{kl}g_{jq}\nonumber \\
 &  & +\frac{1}{4}g_{ip}\tilde{\partial}^{l}\left(g_{lk}\tilde{\partial}^{k}g^{ij}\right)g_{jq}+\frac{1}{4}g_{ip}\partial_{l}\left(g^{lk}\partial_{k}g^{ij}\right)g_{jq}+\frac{1}{4}g_{ip}\tilde{\partial}^{j}g_{kl}\tilde{\partial}^{l}g^{ik}g_{jq}+\frac{1}{4}g_{ip}\tilde{\partial}^{i}g_{kl}\tilde{\partial}^{l}g^{jk}g_{jq}\nonumber \\
 &  & -\frac{1}{4}g_{ip}\tilde{\partial}^{l}\left(g_{kl}\tilde{\partial}^{j}g^{ik}\right)g_{jq}-\frac{1}{4}g_{ip}\tilde{\partial}^{l}\left(g_{kl}\tilde{\partial}^{i}g^{jk}\right)g_{jq}-\frac{1}{4}g_{ip}\partial_{l}\left(g^{ki}\partial_{k}g^{lj}\right)g_{jq}-\frac{1}{4}g_{ip}\partial_{l}\left(g^{kj}\partial_{k}g^{li}\right)g_{jq}\nonumber \\
 &  & +\frac{1}{2}\tilde{\partial}^{i}d\left(g_{ij}\tilde{\partial}^{j}g_{pq}\right)+\frac{1}{2}\partial_{i}d\left(g^{ij}\partial_{j}g_{pq}\right)+2\partial_{p}\partial_{q}d\nonumber \\
 &  & -\frac{1}{2}\tilde{\partial}^{i}d\left(g_{jp}\tilde{\partial}^{j}g_{iq}\right)-\frac{1}{2}\tilde{\partial}^{i}d\left(g_{jq}\tilde{\partial}^{j}g_{ip}\right)-\frac{1}{2}\partial_{i}d\left(g^{ji}\partial_{q}g_{pj}\right)-\frac{1}{2}\partial_{i}d\left(g^{ji}\partial_{p}g_{qj}\right)\nonumber \\
 &  & -\frac{1}{2}g_{ip}\tilde{\partial}^{l}d\left(g_{lk}\tilde{\partial}^{k}\tilde{g}^{ij}\right)g_{jq}-\frac{1}{2}g_{ip}\partial_{l}d\left(g^{lk}\partial_{k}g^{ij}\right)g_{jq}-2g_{ip}\tilde{\partial}^{i}\tilde{\partial}^{j}dg_{jq}\nonumber \\
 &  & +\frac{1}{2}g_{ip}\tilde{\partial}^{l}d\left(g_{kl}\tilde{\partial}^{j}g^{ik}\right)g_{jq}+\frac{1}{2}g_{ip}\tilde{\partial}^{l}d\left(g_{kl}\tilde{\partial}^{i}g^{jk}\right)g_{jq}\nonumber \\
 &  & +\frac{1}{2}g_{ip}\partial_{l}d\left(g^{ki}\partial_{k}g^{lj}\right)g_{jq}+\frac{1}{2}g_{ip}\partial_{l}d\left(g^{kj}\partial_{k}g^{li}\right)g_{jq},\\
\nonumber \\
\mathcal{R}_{\hat{\mathbf{1}}\left(p\right)\hat{\mathbf{2}}\left(q\right)} & = & \frac{1}{8}\tilde{\partial}^{p}n^{2}\partial_{q}n^{-2}+\frac{1}{8}\tilde{\partial}^{p}n^{-2}\partial_{q}n^{2}\nonumber \\
 &  & +\frac{1}{8}\tilde{\partial}^{p}g_{ij}\partial_{q}g^{ij}+\frac{1}{8}\tilde{\partial}^{p}g^{ij}\partial_{q}g_{ij}-\frac{1}{4}\partial_{q}g_{ij}\tilde{\partial}^{j}g^{pi}-\frac{1}{4}\tilde{\partial}^{p}g^{ij}\partial_{j}g_{qi}\nonumber \\
 &  & +\frac{1}{4}\tilde{\partial}^{i}\left(g_{ji}\partial_{q}g^{pj}\right)+\frac{1}{4}\partial_{i}\left(g^{ji}\tilde{\partial}^{p}g_{qj}\right)+\frac{1}{4}\tilde{\partial}^{i}\left(g^{jp}\partial_{j}g_{iq}\right)+\frac{1}{4}\partial_{i}\left(g_{jq}\tilde{\partial}^{j}g^{ip}\right)\nonumber \\
 &  & -\frac{1}{8}g^{ip}\partial_{i}g_{kl}\tilde{\partial}^{j}g^{kl}g_{jq}-\frac{1}{8}g^{ip}\partial_{i}g^{kl}\tilde{\partial}^{j}g_{kl}g_{jq}+\frac{1}{4}g^{ip}\partial_{i}g_{kl}\tilde{\partial}^{l}g^{jk}g_{jq}+\frac{1}{4}g^{ip}\tilde{\partial}^{j}g^{kl}\partial_{l}g_{ik}g_{jq}\nonumber \\
 &  & -\frac{1}{4}g^{ip}\tilde{\partial}^{l}\left(g_{kl}\partial_{i}g^{jk}\right)g_{jq}-\frac{1}{4}g^{ip}\partial_{l}\left(g^{kl}\tilde{\partial}^{j}g_{ik}\right)g_{jq}-\frac{1}{4}g^{ip}\partial_{l}\left(g_{ki}\tilde{\partial}^{k}g^{lj}\right)g_{jq}-\frac{1}{4}g^{ip}\tilde{\partial}^{l}\left(g^{kj}\partial_{k}g_{li}\right)g_{jq}\nonumber \\
 &  & +2\tilde{\partial}^{p}\partial_{q}d-\tilde{\partial}^{i}d\left(g^{jp}\partial_{j}g_{iq}\right)-\partial_{i}d\left(g_{jq}\tilde{\partial}^{j}g^{ip}\right)\nonumber \\
 &  & -\tilde{\partial}^{i}d\left(g_{ji}\partial_{q}g^{pj}\right)-\partial_{i}d\left(g^{ji}\tilde{\partial}^{p}g_{qj}\right)-2g^{ip}\partial_{i}\tilde{\partial}^{j}dg_{jq}\nonumber \\
 &  & +\frac{1}{2}g^{ip}\tilde{\partial}^{l}d\left(g_{kl}\partial_{i}g^{jk}\right)g_{jq}+\frac{1}{2}g^{ip}\partial_{l}d\left(g^{kl}\tilde{\partial}^{j}g_{ik}\right)g_{jq}\nonumber \\
 &  & +\frac{1}{2}g^{ip}\tilde{\partial}^{l}d\left(g^{kj}\partial_{k}g_{li}\right)g_{jq}+\frac{1}{2}g^{ip}\partial_{l}d\left(g_{ki}\tilde{\partial}^{k}g^{lj}\right)g_{jq},\\
\nonumber \\
\mathcal{R}_{\hat{\mathbf{2}}\left(p\right)\hat{\mathbf{1}}\left(q\right)} & = & \frac{1}{8}\partial_{p}n^{2}\tilde{\partial}^{q}n^{-2}+\frac{1}{8}\partial_{p}n^{-2}\tilde{\partial}^{q}n^{2}\nonumber \\
 &  & +\frac{1}{8}\partial_{p}g_{ij}\tilde{\partial}^{q}g^{ij}+\frac{1}{8}\partial_{p}g^{ij}\tilde{\partial}^{q}g_{ij}-\frac{1}{4}\partial_{p}g_{ij}\tilde{\partial}^{j}g^{qi}-\frac{1}{4}\tilde{\partial}^{q}g^{ij}\partial_{j}g_{pi}\nonumber \\
 &  & +\frac{1}{4}\tilde{\partial}^{i}\left(g_{ji}\partial_{p}g^{qj}\right)+\frac{1}{4}\partial_{i}\left(g^{ji}\tilde{\partial}^{q}g_{pj}\right)+\frac{1}{4}\tilde{\partial}^{i}\left(g^{jq}\partial_{j}g_{ip}\right)+\frac{1}{4}\partial_{i}\left(g_{jp}\tilde{\partial}^{j}g^{iq}\right)\nonumber \\
 &  & -\frac{1}{8}g_{ip}\tilde{\partial}^{i}g_{kl}\partial_{j}g^{kl}g^{jq}-\frac{1}{8}g_{ip}\tilde{\partial}^{i}g^{kl}\partial_{j}g_{kl}g^{jq}+\frac{1}{4}g_{ip}\partial_{j}g_{kl}\tilde{\partial}^{l}g^{ik}g^{jq}+\frac{1}{4}g_{ip}\tilde{\partial}^{i}g^{kl}\partial_{l}g_{jk}g^{jq}\nonumber \\
 &  & -\frac{1}{4}g_{ip}\tilde{\partial}^{l}\left(g_{kl}\partial_{j}g^{ik}\right)g^{jq}-\frac{1}{4}g_{ip}\partial_{l}\left(g^{kl}\tilde{\partial}^{i}g_{jk}\right)g^{jq}-\frac{1}{4}g_{ip}\tilde{\partial}^{l}\left(g^{ki}\partial_{k}g_{lj}\right)g^{jq}-\frac{1}{4}g_{ip}\partial_{l}\left(g_{kj}\tilde{\partial}^{k}g^{li}\right)g^{jq}\nonumber \\
 &  & +2\partial_{p}\tilde{\partial}^{q}d-\tilde{\partial}^{i}d\left(g^{jq}\partial_{j}g_{ip}\right)-\partial_{i}d\left(g_{jp}\tilde{\partial}^{j}g^{iq}\right)\nonumber \\
 &  & -\tilde{\partial}^{i}d\left(g_{ji}\partial_{p}g^{qj}\right)-\partial_{i}d\left(g^{ji}\tilde{\partial}^{q}g_{pj}\right)-2g_{ip}\tilde{\partial}^{i}\partial_{j}dg^{jq}\nonumber \\
 &  & +\frac{1}{2}g_{ip}\tilde{\partial}^{l}d\left(g_{kl}\partial_{j}g^{ik}\right)g^{jq}+\frac{1}{2}g_{ip}\partial_{l}d\left(g^{kl}\tilde{\partial}^{i}g_{jk}\right)g^{jq}\nonumber \\
 &  & +\frac{1}{2}g_{ip}\tilde{\partial}^{l}d\left(g^{ki}\partial_{k}g_{lj}\right)g^{jq}+\frac{1}{2}g_{ip}\partial_{l}d\left(g_{kj}\tilde{\partial}^{k}g^{li}\right)g^{jq}.
\end{eqnarray}

\noindent On the other hand, the EOM for the dilaton (\ref{eq:DLA EOM 2})
is expressed as:

\begin{eqnarray}
 &  & \frac{1}{8}g_{ij}\tilde{\partial}^{i}g_{kl}\tilde{\partial}^{j}g^{kl}+\frac{1}{8}g_{ij}\tilde{\partial}^{i}g^{kl}\tilde{\partial}^{j}g_{kl}+\frac{1}{8}g^{ij}\partial_{i}g_{kl}\partial_{j}g^{kl}+\frac{1}{8}g^{ij}\partial_{i}g^{kl}\partial_{j}g_{kl}\nonumber \\
 &  & -\frac{1}{2}g_{ij}\tilde{\partial}^{i}g_{kl}\tilde{\partial}^{k}g^{jl}-\frac{1}{2}g^{ij}\partial_{i}g^{kl}\partial_{k}g_{jl}-\tilde{\partial}^{i}\tilde{\partial}^{j}g_{ij}-\partial_{i}\partial_{j}g^{ij}-4g_{ij}\tilde{\partial}^{i}d\tilde{\partial}^{j}d-4g^{ij}\partial_{i}d\partial_{j}d\nonumber \\
 &  & +4\tilde{\partial}^{i}g_{ij}\tilde{\partial}^{j}d+4\partial_{i}g^{ij}\partial_{j}d+4g_{ij}\tilde{\partial}^{i}\tilde{\partial}^{j}d+4g^{ij}\partial_{i}\partial_{j}d+\frac{1}{4}g_{ij}\tilde{\partial}^{i}n^{2}\tilde{\partial}^{j}n^{-2}+\frac{1}{4}g^{ij}\partial_{i}n^{2}\partial_{j}n^{-2}\nonumber \\
 &  & -\frac{1}{4}n^{-2}\partial_{t}g_{kl}\partial_{t}g^{kl}+\frac{1}{4}n^{-2}\partial_{t}n^{2}\partial_{t}n^{-2}+\partial_{t}\partial_{t}n^{-2}-4\partial_{t}n^{-2}\partial_{t}d=0.
\end{eqnarray}

\noindent Next, we consider the $D=1+\left(2+2\right)$ dimensional
black hole line element:

\begin{equation}
dS^{2}=-n^{2}dt^{2}+\mathcal{H}_{\hat{M}\hat{N}}dX^{\hat{M}}dX^{\hat{N}}=-n\left(r\right)^{2}dt^{2}+g^{ij}d\tilde{x}_{i}d\tilde{x}_{j}+g_{ij}dx^{i}dx^{j},
\end{equation}

\noindent where the strong constraint $\eta_{\hat{M}\hat{N}}\partial^{\hat{M}}\partial^{\hat{N}}\left(\cdots\right)=0$
is imposed, and the metric assumes the form:

\begin{equation}
g_{ij}\left(r\right)=\left(\begin{array}{cc}
B\left(r\right) & 0\\
0 & C\left(r\right)
\end{array}\right).
\end{equation}

\noindent Furthermore, due to symmetries among generalized Ricci scalars
$\mathcal{R}_{M\left(p\right)N\left(q\right)}$,

\begin{eqnarray}
\mathcal{R}_{\hat{\mathbf{1}}\left(p\right)\mathbf{1}\left(q\right)} & \underleftrightarrow{g^{\bullet\bullet}\leftrightarrow g_{\bullet\bullet},\quad\tilde{\partial}^{\bullet}\leftrightarrow\partial_{\bullet}} & \mathcal{R}_{\hat{\mathbf{2}}\left(p\right)\hat{\mathbf{2}}\left(q\right)},\nonumber \\
\mathcal{R}_{\hat{\mathbf{1}}\left(p\right)\hat{\mathbf{2}}\left(q\right)} & \underleftrightarrow{g^{\bullet\bullet}\leftrightarrow g_{\bullet\bullet},\quad\tilde{\partial}^{\bullet}\leftrightarrow\partial_{\bullet}} & \mathcal{R}_{\hat{\mathbf{2}}\left(p\right)\hat{\mathbf{1}}\left(q\right)},
\end{eqnarray}

\noindent we only need to calculate one set of them. Here, we compute
$\mathcal{R}_{\hat{\mathbf{2}}\left(p\right)\hat{\mathbf{2}}\left(q\right)}$
and $\mathcal{R}_{\hat{\mathbf{2}}\left(p\right)\hat{\mathbf{1}}\left(q\right)}$.
Consequently, the gravitational EOM (\ref{eq:GR EOM 2}) become:

\begin{eqnarray}
\mathcal{R}_{tt} & = & nB\left[\left(\frac{\tilde{\partial}^{r}B}{B}-\frac{\tilde{\partial}^{r}n}{n}\right)\tilde{\partial}^{r}n+\tilde{\partial}^{r}\tilde{\partial}^{r}n-2\tilde{\partial}^{r}d\tilde{\partial}^{r}n\right]\nonumber \\
 &  & -n\frac{1}{B}\left[\left(\frac{\partial_{r}B}{B}+\frac{\partial_{r}n}{n}\right)\partial_{r}n-\partial_{r}\partial_{r}n+2\partial_{r}d\partial_{r}n\right]=0,\nonumber \\
\nonumber \\
\mathcal{R}_{\hat{\mathbf{2}}\left(r\right)\hat{\mathbf{2}}\left(r\right)} & = & B\left[B\left(\frac{\tilde{\partial}^{r}n}{n}\right)^{2}+\frac{1}{4}B\left(\frac{\tilde{\partial}^{r}C}{C}\right)^{2}+\frac{1}{2}\tilde{\partial}^{r}\tilde{\partial}^{r}B-\frac{1}{4}\frac{\left(\tilde{\partial}^{r}B\right)^{2}}{B}-\tilde{\partial}^{r}d\tilde{\partial}^{r}B-2B\tilde{\partial}^{r}\tilde{\partial}^{r}d\right]\nonumber \\
 &  & -\frac{1}{B}\left[B\left(\frac{\partial_{r}n}{n}\right)^{2}+\frac{1}{4}B\left(\frac{\partial_{r}C}{C}\right)^{2}-\frac{1}{2}\partial_{r}\partial_{r}B+\frac{3}{4}\frac{\left(\partial_{r}B\right)^{2}}{B}+\partial_{r}d\partial_{r}B-2B\partial_{r}\partial_{r}d\right]=0,\nonumber \\
\nonumber \\
\mathcal{R}_{\hat{\mathbf{2}}\left(\varphi\right)\hat{\mathbf{2}}\left(\varphi\right)} & = & \frac{1}{2}B\left[-\frac{\tilde{\partial}^{r}B}{B}\tilde{\partial}^{r}C-\tilde{\partial}^{r}\tilde{\partial}^{r}C+\frac{\left(\tilde{\partial}^{r}C\right)^{2}}{C}+2\tilde{\partial}^{r}d\tilde{\partial}^{r}C\right]\nonumber \\
 &  & -\frac{1}{2}\frac{1}{B}\left[-\frac{\partial_{r}B}{B}\partial_{r}C+\partial_{r}\partial_{r}C-\frac{\left(\partial_{r}C\right)^{2}}{C}-2\partial_{r}d\partial_{r}C\right]=0.\label{eq:R eom}
\end{eqnarray}

\noindent Moreover, to compute the black hole solution in non-critical
string theory, we must incorporate the cosmological constant into
the action. This constant does not break the $O\left(d,d\right)$
symmetry and the diffeomorphism of the action:

\begin{equation}
S_{\mathrm{m}}=\int dt\int d^{D-1}xd^{D-1}\tilde{x}e^{-2d}\lambda^{2}.
\end{equation}

\noindent Therefore, the EOM for the dilaton (\ref{eq:DLA EOM 2})
is expressed as:

\begin{eqnarray}
B\left[-\left(\frac{\tilde{\partial}^{r}n}{n}\right)^{2}+\frac{1}{4}\left(\frac{\tilde{\partial}^{r}B}{B}\right)^{2}-\frac{1}{4}\left(\frac{\tilde{\partial}^{r}C}{C}\right)^{2}-\frac{\tilde{\partial}^{r}\tilde{\partial}^{r}B}{B}+4\frac{\tilde{\partial}^{r}B}{B}\tilde{\partial}^{r}d-4\tilde{\partial}^{r}d\tilde{\partial}^{r}d+4\tilde{\partial}^{r}\tilde{\partial}^{r}d\right]\nonumber \\
+\frac{1}{B}\left[-\left(\frac{\partial_{r}n}{n}\right)^{2}-\frac{7}{4}\left(\frac{\partial_{r}B}{B}\right)^{2}-\frac{1}{4}\left(\frac{\partial_{r}C}{C}\right)^{2}+\frac{\partial_{r}\partial_{r}B}{B}-4\frac{\partial_{r}B}{B}\partial_{r}d-4\partial_{r}d\partial_{r}d+4\partial_{r}\partial_{r}d\right] & = & -\lambda^{2}.\nonumber \\
\label{eq:dilaton eom}
\end{eqnarray}

\noindent At first, we attempt to derive a solution corresponding
to the 3rd kind of solution found in standard DFT as discussed in
the previous section. The solution, which violates constraints, is
given by:

\begin{equation}
ds^{2}=\left[-n\left(r,\tilde{r}\right)^{2}dt^{2}+dr^{2}+dx^{2}\right]+\left[d\tilde{r}^{2}+d\tilde{x}^{2}\right],
\end{equation}

\noindent where

\begin{eqnarray}
n\left(r,\tilde{r}\right) & = & \left(\tanh\left(\frac{\sqrt{2}\lambda}{4}\tilde{r}\right)\right)^{\mp1}\left(\tanh\left(\frac{\sqrt{2}\lambda}{4}r\right)\right)^{\pm1},\nonumber \\
d\left(r,\tilde{r}\right) & = & -\frac{1}{2}\log\left(\sinh\left(\frac{\sqrt{2}\lambda}{2}\tilde{r}\right)\sinh\left(\frac{\sqrt{2}\lambda}{2}r\right)\right).
\end{eqnarray}

\noindent This configuration represents a double black string solution.
However, it violates the strong constraint due to its dependence on
dual coordinates $\left(r,\tilde{r}\right)$. Further investigation
into this solution is postponed until a compelling reason to relax
the strong constraint and interpret dual coordinates emerges. Alternatively,
we impose the strong constraint and derive two types of solutions
for the EOM (\ref{eq:R eom}) and (\ref{eq:dilaton eom}), corresponding
to the 1st kind of solution discussed in the previous section. The
solution is

\begin{equation}
ds^{2}=\left[-n\left(r\right)^{2}dt^{2}+B\left(r\right)dr^{2}+C\left(r\right)d\varphi^{2}\right]+\left[B\left(r\right)^{-1}d\tilde{r}^{2}+C\left(r\right)^{-1}d\tilde{\varphi}^{2}\right],
\end{equation}

\noindent where

\begin{eqnarray}
 &  & n\left(r\right)=\left(\sqrt{\left(1-\frac{M}{r}\right)^{a}}\right)^{\pm1},\qquad B\left(r\right)=\left(1-\frac{M}{r}\right)^{-c}\frac{1}{\lambda^{2}r^{2}},\qquad C\left(r\right)=\left(\left(1-\frac{M}{r}\right)^{b}\frac{1}{\lambda^{2}}\right)^{\pm1},\nonumber \\
 &  & d=\mathrm{const},\qquad a^{2}+b^{2}=1,\qquad c=1.
\end{eqnarray}

\noindent The choices of $\pm$ for $n\left(r\right)$ and $C\left(r\right)$
do not need to be the same simultaneously. The dual solution is given
by:

\begin{equation}
ds^{2}=\left[-n\left(\tilde{r}\right)^{2}dt^{2}+B\left(\tilde{r}\right)^{-1}d\tilde{r}^{2}+C\left(\tilde{r}\right)^{-1}d\tilde{\varphi}^{2}\right]+\left[B\left(\tilde{r}\right)dr^{2}+C\left(\tilde{r}\right)d\varphi^{2}\right]
\end{equation}

\noindent where

\begin{eqnarray}
 &  & n\left(\tilde{r}\right)=\left(\sqrt{\left(1-\frac{M}{\tilde{r}}\right)^{a}}\right)^{\mp1},\qquad B\left(\tilde{r}\right)=\left(1-\frac{M}{\tilde{r}}\right)^{c}\lambda^{2}\tilde{r}^{2},\qquad C\left(\tilde{r}\right)=\left(\left(1-\frac{M}{\tilde{r}}\right)^{-b}\lambda^{2}\right)^{\mp1},\nonumber \\
 &  & d=\mathrm{const},\qquad a^{2}+b^{2}=1,\qquad c=1.
\end{eqnarray}

\noindent A similar solution in the low-energy effective action was
studied in reference \cite{Cadoni:1996ax}. Note that our solutions
apply to both compact and noncompact $\varphi$.

For simplicity, we analyze only one of these solutions, specifically:

\begin{eqnarray}
ds^{2} & = & \left[-\left(1-\frac{M}{r}\right)^{a}dt^{2}+\left(1-\frac{M}{r}\right)^{-c}\frac{1}{\lambda^{2}r^{2}}dr^{2}+\left(1-\frac{M}{r}\right)^{b}\frac{1}{\lambda^{2}}d\varphi^{2}\right]\nonumber \\
 &  & +\left[\left(1-\frac{M}{r}\right)^{c}\lambda^{2}r^{2}d\tilde{r}^{2}+\left(1-\frac{M}{r}\right)^{-b}\lambda^{2}d\tilde{\varphi}^{2}\right].\label{eq:final solution}
\end{eqnarray}

\noindent To determine its physical interpretation, we introduce two
key results:

\subsubsection*{A. Generalized scalar curvature}

\noindent To investigate the singularities of this metric, we need
to consider the scalar curvature within the framework of this theory.
The geometry of DFT is structured around generalized diffeomorphisms.
Therefore, the connection must incorporate covariant derivatives that
respect these symmetries. This connection can be defined analogously
to the Levi-Civita connection in Riemannian geometry, ensuring compatibility
with the generalized metric and the $O\left(d,d\right)$ structure.
Using this connection, the generalized Ricci curvature is derived
under the condition of vanishing torsion. However, the generalized
Ricci scalar $\mathcal{R}$ and tensor $\mathcal{R}_{MN}$ are constrained
to vanish due to satisfying the equations of motion for the graviton
and dilaton: $\mathcal{R}=0$ and $\mathcal{R}_{MN}=0$. This property
complicates the measurement of curvature, even in spacetime singularities
within DFT \cite{Arvanitakis:2016zes}. In our solutions, incorporating
the cosmological constant in DFT leads to the generalized Ricci scalar
being

\begin{equation}
\mathcal{R}=-\lambda^{2},
\end{equation}

\noindent indicating that our solutions represent the generalized
AdS vacuum solutions in double coordinates. This additional term is
valid as the DFT action should reduce to the non-critical low-energy
effective action after imposing the strong constraint \cite{Wu:2013sha,Lv:2014ava}.
Importantly, DFT possesses the generalized AdS vacuum solutions due
to the generalized Ricci scalar includes dilatonic effects. However,
we wish to note that this solution differs from the traditional AdS
vacuum, as it is derived from the generalized Ricci scalar, which
is defined on the doubled spacetime. Further studies are needed to
fully understand its physical implications.

\subsubsection*{B. Reduction to low-energy effective action}

To impose the strong constraint $\tilde{\partial}\left(\ldots\right)=0$,
requiring all fields to be independent of the dual coordinates, the
DFT action simplifies to the traditional low-energy effective action:

\begin{equation}
\left.S_{\mathrm{DFT}}\right|_{\tilde{\partial}=0}=\int dt\int d^{D-1}xd^{D-1}\tilde{x}\sqrt{-g}e^{-2\phi}\left[R+4\left(\nabla\phi\right)^{2}-\frac{1}{12}H_{\mu\nu\rho}H^{\mu\nu\rho}\right].
\end{equation}

\noindent Alternatively, this also can be expressed as:

\begin{equation}
\left.\mathcal{R}\right|_{\tilde{\partial}=0}=\sqrt{-g}e^{-2\phi}\left[R+4\left(\nabla\phi\right)^{2}-\frac{1}{12}H_{\mu\nu\rho}H^{\mu\nu\rho}\right].
\end{equation}

\noindent While integrating out the dual coordinates $\tilde{x}$
only yields an overall constant in the action, it may seem negligible.
However, to retain the novel insights of DFT, we preserve this term
and interpret the generalized metric as the complete metric of spacetime.
Moreover, this reduction implies that the undoubled geometry described
by $x$ can be entirely governed by the traditional low-energy effective
action, without the need for any additional theories. This indicates
that the traditional interpretation of black holes in the low-energy
effective action also applies to DFT black holes. Thus, the Kretschmann
scalar for the undoubled spacetime (spanned by $x$) of the generalized
metric (\ref{eq:final solution}) is given by

\begin{equation}
R_{\mu\nu\rho\sigma}R^{\mu\nu\rho\sigma}=\frac{\lambda^{4}M^{2}\left(1-\frac{M}{r}\right)^{2c}\left(a^{4}M^{2}+2a^{3}M\left(cM-2r\right)+a^{2}\left(b^{2}M^{2}+\left(cM-2r\right)^{2}\right)+b^{2}\left(bM+cM-2r\right)^{2}\right)}{4\left(M-r\right)^{4}}.
\end{equation}

\vspace*{4.0ex}

Based on these results, we aim to understand the physical significance
of the black hole solution (\ref{eq:final solution}) in DFT. In the
framework of DFT, the entire doubled metric of spacetime is given
by:

\begin{equation}
ds^{2}=\underset{\mathrm{AdS\;vacuum\;in\;DFT}}{\underbrace{\underset{\mathrm{Black\;hole\;in\;ordinary\;string\;theory}}{\underbrace{{\color{blue}\left[-n\left(r\right)^{2}dt^{2}+B\left(r\right)dr^{2}+C\left(r\right)d\varphi^{2}\right]}}}+{\color{cyan}\left[B\left(r\right)^{-1}d\tilde{r}^{2}+C\left(r\right)^{-1}d\tilde{\varphi}^{2}\right]}}},
\end{equation}

\begin{figure}[H]
\begin{centering}

\tikzset{every picture/.style={line width=0.75pt}} %set default line width to 0.75pt

\begin{tikzpicture}[x=0.75pt,y=0.75pt,yscale=-1,xscale=1]

%Shape: Ellipse [id:dp430437141477477]
\draw  [fill={rgb, 255:red, 80; green, 227; blue, 194 }  ,fill opacity=1 ] (88,144.5) .. controls (88,104.46) and (192.09,72) .. (320.5,72) .. controls (448.91,72) and (553,104.46) .. (553,144.5) .. controls (553,184.54) and (448.91,217) .. (320.5,217) .. controls (192.09,217) and (88,184.54) .. (88,144.5) -- cycle ;
%Shape: Ellipse [id:dp5077383389130146]
\draw  [fill={rgb, 255:red, 74; green, 144; blue, 226 }  ,fill opacity=1 ] (129,146.5) .. controls (129,120.82) and (169.74,100) .. (220,100) .. controls (270.26,100) and (311,120.82) .. (311,146.5) .. controls (311,172.18) and (270.26,193) .. (220,193) .. controls (169.74,193) and (129,172.18) .. (129,146.5) -- cycle ;
%Straight Lines [id:da5413584672906293]
\draw    (92,215) -- (161.33,169.1) ;
\draw [shift={(163,168)}, rotate = 146.5] [color={rgb, 255:red, 0; green, 0; blue, 0 }  ][line width=0.75]    (10.93,-3.29) .. controls (6.95,-1.4) and (3.31,-0.3) .. (0,0) .. controls (3.31,0.3) and (6.95,1.4) .. (10.93,3.29)   ;

% Text Node
\draw (335,127) node [anchor=north west][inner sep=0.75pt]   [align=left] {\begin{minipage}[lt]{126.68pt}\setlength\topsep{0pt}
\textbf{Doubled spacetime }$\displaystyle \left( x,\tilde{x}\right)$
\begin{center}
\textbf{AdS vacuum}
\end{center}

\end{minipage}};
% Text Node
\draw (140,130) node [anchor=north west][inner sep=0.75pt]   [align=left] {\begin{minipage}[lt]{114.25pt}\setlength\topsep{0pt}
\textbf{Ordinary spacetime }$\displaystyle ( x)$
\begin{center}
\textbf{Black holes}
\end{center}

\end{minipage}};
% Text Node
\draw (16,228) node [anchor=north west][inner sep=0.75pt]   [align=left] {\begin{minipage}[lt]{139.75pt}\setlength\topsep{0pt}
\begin{center}
completely described by\\the low energy effective action
\end{center}

\end{minipage}};

\end{tikzpicture}
\par\end{centering}
\centering{}\caption{\label{fig:DFT solution}The dark blue region represents the conventional
undoubled spacetime described by the low-energy effective action,
which contains the black holes. When the additional region from the
double coordinates is included, only DFT can describe the entire doubled
spacetime, consistently characterizing it as the generalized AdS vacuum.}
\end{figure}
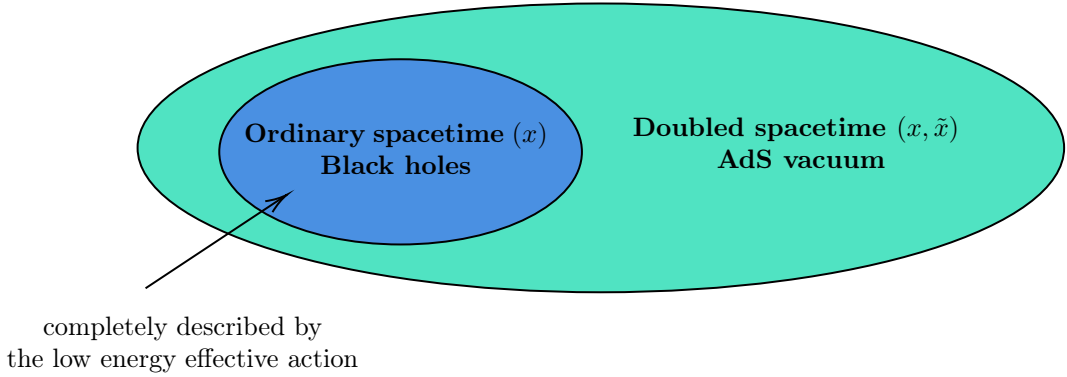

\noindent This entire metric describes the generalized AdS vacuum
in double coordinates. The first bracket corresponds to the black
hole solution known from ordinary string theory, while the second
bracket represents the additional dimensions introduced by DFT. Since
the first bracket is fully determined by the low-energy effective
action, the usual interpretations in this framework can be directly
applied. This suggests that this conventional, undoubled spacetime
$x$ is self-contained and exhibits its own dynamics. If we consider
the additional part $\left(\tilde{r},\tilde{\varphi}\right)$, there
is no action to describe its geometry independently. Therefore, we
must consider the entire doubled spacetime. DFT governs its dynamics,
and it manifests as the generalized AdS vacuum. In other words, if
we obscure the doubled coordinates and focus only on the conventional
region, the black hole and its singularity emerge as solutions of
the low-energy effective action, as depicted in Figure. (\ref{fig:DFT solution}).

\vspace*{2.0ex}

To specify, we examine the following double black string solutions
with specific values for $a$, $b$ and $c$:

\paragraph*{1. When $a=1$, $b=0$,\textcolor{brown}{{} }$c=1$, we obtain:}

\begin{equation}
ds^{2}=\left[-\left(1-\frac{M}{r}\right)dt^{2}+\frac{1}{r\left(r-M\right)}\frac{1}{\lambda^{2}}dr^{2}+\frac{1}{\lambda^{2}}d\varphi^{2}\right]+\left[r\left(r-M\right)\lambda^{2}d\tilde{r}^{2}+\lambda^{2}d\tilde{\varphi}^{2}\right]
\end{equation}

\noindent If we only consider the first square bracket, the corresponding
Kretschmann scalar is $R_{\mu\nu\rho\sigma}R^{\mu\nu\rho\sigma}=\frac{\lambda^{4}M^{2}}{r^{2}}$,
indicating this black hole solution possesses a spacetime singularity:
$r=0$ where the event horizon is located at $r=M$. Considering the
whole metric, there are two time-like directions $dr^{2}$ and $d\tilde{r}^{2}$
acrossing the event horizon. Additionally, accounting for the coordinates
$\left(\tilde{r},\tilde{\varphi}\right)$, there is a coordinate singularity
as $r\rightarrow\infty$. However, it is impossible to measure the
curvature in the region $\left(\tilde{r},\tilde{\varphi}\right)$
independently. When considering the doubled coordinates, the metric
corresponds to the generalized AdS vacuum.

\vspace*{2.0ex}

\paragraph*{2. When $a=-1$, $b=0$,\textcolor{brown}{{} }$c=1$, we have:}

\begin{equation}
ds^{2}=\left[-\left(1-\frac{M}{r}\right)^{-1}dt^{2}+\frac{1}{r\left(r-M\right)}\frac{1}{\lambda^{2}}dr^{2}+\frac{1}{\lambda^{2}}d\varphi^{2}\right]+\left[r\left(r-M\right)\lambda^{2}d\tilde{r}^{2}+\lambda^{2}d\tilde{\varphi}^{2}\right],\qquad M\leq0,
\end{equation}

\noindent Considering the first square bracket, it is the solution
of low-energy effective action and the corresponding Kretschmann scalar
is $R_{\mu\nu\rho\sigma}R^{\mu\nu\rho\sigma}=\frac{\lambda^{4}M^{2}}{\left(M-r\right)^{2}}$.
This solution has a single curvature singularity at $r=M$, with a
coordinate singularity as $r\rightarrow\infty$. And for complete
metric, the metric still corresponds to the generalized AdS vacuum.

\vspace*{2.0ex}

\paragraph*{3. When $a=0$, $b=\pm1$,\textcolor{brown}{{} }$c=1$, we have:}

\begin{eqnarray}
ds^{2} & = & \left[-dt^{2}+\left(1-\frac{M}{r}\right)^{-1}\frac{1}{\lambda^{2}r^{2}}dr^{2}+\left(1-\frac{M}{r}\right)^{\pm1}\frac{1}{\lambda^{2}}d\varphi^{2}\right]\nonumber \\
 &  & +\left[\left(1-\frac{M}{r}\right)\lambda^{2}r^{2}d\tilde{r}^{2}+\left(1-\frac{M}{r}\right)^{\mp1}\lambda^{2}d\tilde{\varphi}^{2}\right],\qquad M\geq0,
\end{eqnarray}

\noindent Here, $b=1$ for $M\geq0$ and $b=-1$ for $M\leq0$. When
$b=-1$, there is a conical singularity at $r=M$ due to $R_{\mu\nu\rho\sigma}R^{\mu\nu\rho\sigma}=\frac{\lambda^{4}M^{2}}{\left(M-r\right)^{2}}$.
When $b=1$, the curvature singularity is at $r=0$ because $R_{\mu\nu\rho\sigma}R^{\mu\nu\rho\sigma}=\frac{\lambda^{4}M^{2}}{r^{2}}$.
Furthermore, when the doubled coordinates are considered, the metric
corresponds to the generalized AdS vacuum.

\vspace*{2.0ex}

\paragraph*{4. When $a=b=c=0$, the solution is given as follows:}

\begin{eqnarray}
ds^{2} & = & \left[-dt^{2}+\frac{1}{\lambda^{2}r^{2}}dr^{2}+\lambda^{\pm2}d\varphi^{2}\right]+\left[\lambda^{2}r^{2}d\tilde{r}^{2}+\lambda^{\mp2}d\tilde{\varphi}^{2}\right].
\end{eqnarray}

\noindent The first square bracket of this metric describe flat spacetime,
as $R_{\mu\nu\rho\sigma}R^{\mu\nu\rho\sigma}=0$. This implies that
the half-geometry of the generalized AdS vacuum in the doubled geometry
can also manifest as flat spacetime when the doubled coordinates are
ignored.

\vspace*{4.0ex}

\paragraph*{Comment on the Cosmological Solutions}

Finally, we wish to note that our calculation methods
for DFT black holes are also applicable to DFT cosmology. We start
with the low-energy effective action, assuming a vanishing Kalb-Ramond
field:

\begin{equation}
S=\int d^{D}x\sqrt{-g}e^{-2\phi}\left[R+4\left(\nabla\phi\right)^{2}\right].\label{eq:cos action}
\end{equation}

\noindent The cosmological solutions of DFT can be
constructed through the solutions of this action (\ref{eq:cos action}):

\vspace*{2.0ex}

\noindent \textbf{1st kind of solutions:}{}
If $g_{ij}\left(t\right)$ and $\phi\left(t\right)$ are solutions
of the effective action (\ref{eq:cos action}), the solutions of DFT
that satisfy the strong constraint can be given as:

\noindent 
\begin{equation}
ds^{2}=g^{ij}\left(t\right)d\tilde{x}_{i}d\tilde{x}_{j}+g_{ij}\left(t\right)dx^{i}dx^{j},\qquad or\qquad ds^{2}=\left(g^{-1}\right)^{ij}\left(\tilde{t}\right)d\tilde{x}_{i}d\tilde{x}_{j}+\left(g^{-1}\right)_{ij}\left(\tilde{t}\right)dx^{i}dx^{j}.
\end{equation}

\noindent Note this case works for any cosmological
solution of (\ref{eq:cos action}).

\vspace*{1.0ex}

\noindent \textbf{2nd kind of solutions: }If
$g_{ij}\left(t\right)$ and $\left(g^{-1}\right)_{ij}\left(t\right)$
are $O\left(d,d\right)$ dual solutions of the effective action (\ref{eq:cos action}),
the constraint-satisfying solutions of DFT can be given as:

\begin{equation}
ds^{2}=g^{ij}\left(t\right)d\tilde{x}_{i}d\tilde{x}_{j}+g_{ij}\left(t\right)dx^{i}dx^{j},\qquad or\qquad ds^{2}=\left(g^{-1}\right)^{ij}\left(t\right)d\tilde{x}_{i}d\tilde{x}_{j}+\left(g^{-1}\right)_{ij}\left(t\right)dx^{i}dx^{j},
\end{equation}

\noindent or

\begin{equation}
ds^{2}=g^{ij}\left(\tilde{t}\right)d\tilde{x}_{i}d\tilde{x}_{j}+g_{ij}\left(\tilde{t}\right)dx^{i}dx^{j},\qquad or\qquad ds^{2}=\left(g^{-1}\right)^{ij}\left(\tilde{t}\right)d\tilde{x}_{i}d\tilde{x}_{j}+\left(g^{-1}\right)_{ij}\left(\tilde{t}\right)dx^{i}dx^{j}.
\end{equation}

\noindent For example, the simplest solutions are
given by:

\vspace*{1.0ex}

\noindent 
\begin{equation}
ds^{2}=-d\tilde{t}^{2}+\left|t\right|^{\pm1/\sqrt{D-1}}\delta_{ij}d\tilde{x}_{i}d\tilde{x}_{j}-dt^{2}+\left|t\right|^{\mp1/\sqrt{D-1}}\delta_{ij}dx^{i}dx^{j},\qquad d\left(t\right)=-\frac{1}{2}\ln\left|t\right|,
\end{equation}

\noindent or

\noindent 
\begin{equation}
ds^{2}=-d\tilde{t}^{2}+\left|\tilde{t}\right|^{\pm1/\sqrt{D-1}}\delta_{ij}d\tilde{x}_{i}d\tilde{x}_{j}-dt^{2}+\left|\tilde{t}\right|^{\mp1/\sqrt{D-1}}\delta_{ij}dx^{i}dx^{j},\qquad d\left(\tilde{t}\right)=-\frac{1}{2}\ln\left|\tilde{t}\right|,
\end{equation}

\noindent where $d\left(t\right)$ and $d\left(\tilde{t}\right)$
are $O\left(d,d\right)$ dilaton.

\vspace*{2.0ex}

\noindent \textbf{3rd kind of solutions: }If
$g_{ij}\left(t\right)$ and $\left(g^{-1}\right)_{ij}\left(t\right)$
are solutions of the effective action (\ref{eq:cos action}), we can
construct constraint-violating solutions of DFT such that $g\left(\tilde{t},t\right)=g\left(\tilde{t}\right)\times\left(g^{-1}\right)\left(t\right)$.
These solutions depend on double time coordinates and take the forms:

\begin{equation}
ds^{2}=g^{ij}\left(t,\tilde{t}\right)d\tilde{x}_{i}d\tilde{x}_{j}+g_{ij}\left(t,\tilde{t}\right)dx^{i}dx^{j},\qquad or\qquad ds^{2}=\left(g^{-1}\right)^{ij}\left(t,\tilde{t}\right)d\tilde{x}_{i}d\tilde{x}_{j}+\left(g^{-1}\right)_{ij}\left(t,\tilde{t}\right)dx^{i}dx^{j}.
\end{equation}

\noindent For example, the simplest solutions are
given by:

\vspace*{1.0ex}

\noindent 
\begin{equation}
ds^{2}=-d\tilde{t}^{2}+\left|\frac{t}{\tilde{t}}\right|^{\pm1/\sqrt{D-1}}\delta_{ij}d\tilde{x}_{i}d\tilde{x}_{j}-dt^{2}+\left|\frac{t}{\tilde{t}}\right|^{\mp1/\sqrt{D-1}}\delta_{ij}dx^{i}dx^{j},\qquad d\left(t,\tilde{t}\right)=-\frac{1}{2}\ln\left|tt\right|,
\end{equation}

\noindent or

\noindent 
\begin{equation}
ds^{2}=-d\tilde{t}^{2}+\left|t\tilde{t}\right|^{\pm1/\sqrt{D-1}}\delta_{ij}d\tilde{x}_{i}d\tilde{x}_{j}-dt^{2}+\left|t\tilde{t}\right|^{\mp1/\sqrt{D-1}}\delta_{ij}dx^{i}dx^{j},\qquad d\left(t,\tilde{t}\right)=-\frac{1}{2}\ln\left|t\tilde{t}\right|.
\end{equation}

\section{Conclusion and discussion}

In this paper, we studied black hole solutions in DFT across two main
sections. Initially, we directly calculated black hole solutions within
standard DFT, where metrics involve two time-like directions. These
solutions were categorized into three classes: two imposing to the
strong constraint and one violating it by depending on dual coordinates.
Due to the ambiguity in interpreting solutions with two time-like
directions, further investigation is warranted for these cases. In
the second part, we explored the space/time split DFT, where the theory
only depends on conventional time-like directions, manifesting in
a $1+2\left(D-1\right)$ dimensional spacetime $\left(t,x,\tilde{x}\right)$.
In this framework, we investigated non-critical string theory incorporating
a cosmological constant, also deriving solutions that violate the
constraint. These solutions, similarly, are left for future investigation.
Furthermore, we derived double black string solutions satisfying the
strong constraint. In the context of the entire doubled spacetime,
the generalized Ricci scalar consistently manifests as a negative
constant, indicating the generalized AdS vacuum solutions. Upon hidden
the dual coordinates $x$, the standard geometry described by $\tilde{x}$
reveals black hole solutions and curvature singularities.

Finally, we present some valuable study points for the near future:

\subsubsection*{Cosmology}

Using our method, the cosmology of DFT can be revisited. By incorporating
the cosmological constant, we can derive cosmological solutions. The
corresponding generalized FLRW line element is given by:

\begin{equation}
ds^{2}=-dt^{2}+a^{2}\left(t\right)dx^{i}dx_{i}+a^{-2}\left(t\right)d\tilde{x}^{i}d\tilde{x}_{i}.
\end{equation}

\noindent This solution addresses the double time problem discussed
in reference \cite{Wu:2013sha}. Similar to the black hole solution,
when considering the entire doubled spacetime, the generalized curvature
implies that this solution corresponds to a dS vacuum.

\subsubsection*{Black hole soluions in Type II DFT}

Considering the DFT action of type II theories \cite{Hohm:2011zr,Hohm:2011dv,Hohm:2011cp}:

\begin{equation}
S=\int dxd\tilde{x}\left(e^{-2d}\mathcal{R}\left(\mathcal{H},d\right)+\frac{1}{4}\left(\cancel{\partial}\chi\right)^{\dagger}\mathbb{S}\cancel{\partial}\chi\right),
\end{equation}

\noindent where the first term represents the standard DFT action
describing the NS-NS sector, and the second term is the DFT action
for the massless RR fields. The RR sector introduces two independent
fields. The first field is $\mathbb{S}=\mathbb{S}^{\dagger}$, $\mathbb{S}\in\mathrm{Spin}\left(10,10\right)$.
The second field is a spinor $\chi$, which can be represented by
a Majorana-Weyl spinor of $O\left(10,10\right)$. Using our generating
method, we can also easily obtain the double black $p$-brane solution
of this action. It would be intriguing to explore whether this approach
introduces new insights into the AdS/CFT correspondence.

\noindent \bigskip

\vspace{5mm}

\noindent {\bf Acknowledgements}
This work was supported by NSFC Grant No.12105031, No.12347101 and PSFC Grant No. cstc2021jcyj-bshX0227.

\end{document}